\documentclass[draft,tightenlines,nofootinbib,preprint,aps,eqsecnum]{revtex4}%
\usepackage{amsmath}
\usepackage{amsfonts}
\usepackage{amssymb}
\usepackage{graphicx}%
\providecommand{\U}[1]{\protect\rule{.1in}{.1in}}
\begin{document}
\title{ \ {\huge Baryon Spectrum Analysis using Dirac's Covariant Constraint
Dynamics}}
\author{Joshua F.\ Whitney and Horace W. Crater}

\begin{abstract}
We present a relativistic quark model for the baryons that combines three
related relativistic formalisms. The three-body constraint formalism of
Sazdjian is used to recast three relativistic two-body equations for the three
pairs of interacting quarks into a single relativistically covariant three
body equation for the bound state energies, having a Schrodinger-like
structure. The two-body equations are the Two Body Dirac equations (TBDE) of
constraint dynamics derived by Crater and Van Alstine for combined world
vector and scalar interactions providing the necessary spin dependent and spin
independent interaction terms. The minimal quasipotential formalism of Todorov
is used to provide an invariant framework for the vector and scalar dynamics
used in the TBDE into which is inserted a local simplified version of the
Richardson potential. The spectral results are analyzed and compared to
experiment using a best fit method and several different algorithms, including
a gradient approach, and Monte Carlo method.

\end{abstract}
\maketitle

\today

\vfill\eject

\section{Introduction}

Recent quark model calculations done by Crater et al.
\cite{jim,wongyoon,alstine2009} using covariant Two-Body Dirac equations
(TBDE) in a relativistic constraint dynamics formalism have given a good
description of the meson masses for both light and heavy quarks. \ The good
quality of the fit has been attributed to the exact two-body kinematics merged
with a QCD interaction potential based on vector and scalar potentials that
uses a minimal number of variable parameters. \ The vector potential, in turn,
has a structure originally derived from the classical electrodynamics of
Wheeler and Feynman.\cite{18} \ This structure can also be obtained from
quantum electrodynamics by using a covariant three dimensional truncation of
the Bethe-Salpeter equation \cite{bse} based on the Todorov quasipotential
approach\cite{quasi}, which is then compared to the TBDE.\cite{becker} The
comparison is done in order to identify the appropriate invariant potential
functions that will be used in the potential model. These non-perturbative
(numerical) results hold up well when compared to other methods for meson
spectroscopy. \ 

In this paper, we extend the Hamiltonian constraint dynamics formalism for the
two-body system to the three-body quark problem for baryon
spectroscopy\cite{whitney1}. \ In taking the two-body equations to a
three-body system we still regard the system as the naive quark model in that
the interactions are between each pair of quarks and there is no over-arching
three-body interaction to be considered and no consideration of the effects of
baryon decays on their masses, but the system now has three sets of
interactions instead of just one. \ Thus, all of our interactions are still
two-body interactions, but for three sets of quarks. \ 

The\ Hamiltonian Constraint Dynamics formalism \cite{cnstr} allows for a
relativistic method of accounting for two-body effects. \ In addition,
Constraint Dynamics as developed by Crater and Van Alstine \cite{cra82}
provides not only the usual spin interaction dependence seen in the Dirac
equation but also additional terms needed to make the approach mathematically
consistent. \ It is useful in both a classical and a quantum mechanical
formalism as well as correctly accounting for fine and hyperfine structures in
positronium and muonium systems\cite{exct},\cite{becker}. We review the
two-body formalism with an eye toward its adaptation to the three body system
using Sazdjian's approach to relativistic $N-$body problem\cite{5}

In our adaptation of the two-body formalism to the numerical solution of three
quark bound states, the variational principle\ is used together with a new
type of Gaussian basis wavefunctions of total $JM$ to solve our eigenvalue
equation. \ We show how this new Gaussian basis is used in conjunction with a
variational approach in obtaining the appropriate eigenvalues with the matrix
being truncated after a reasonable limit is reached (in theory an infinitely
large variational matrix would give the exact reflection of the states of the
model). \ Since the effective potentials are dependent on the center of
momentum (c.m.) total energy eigenvalue $w=m_{1}+m_{2}+m_{3}+E$, the standard
approach to using the variational principle must be modified to include a
recursion algorithm with an embedded $E$ dependence in the matrix elements of
the effective Hamiltonian $\mathcal{H}.$ This $E$ will change as we approach
convergence. \ The program then is designed to iteratively solve these
equations until a desired level of convergence is reached. \ The matix
elements of our Hamiltonian can be determined exactly for the kinematics but
including the interacting potentials requires, of course, a numerical
treatment. \ Also as in the two-body case, the interacting potentials we used
\cite{wongyoon} depend only on three parameters characterizing just two
invariant functions, embodying the world vector and scalar potentials
appearing in the TBDE . \ The numerical fitting routine uses a chi-squared
minimization Monte Carlo\ routine combined with a simplified gradient approach
to acquire a best fit for the spectrum of known baryons. \ We have compared
our numerical results to both experimental data and to other theories, most
notably the approach of Capstick and Isgur\cite{2}, while also comparing the
quark masses and potential parameters we obtained in our fit with those found
in two-body the meson spectral results of \cite{wongyoon} \ 

\bigskip

\section{Review of Relativistic Two-Body Constraint Approach}

In this and in the following subsections we present a review of the constraint
formalism in preparation for its implementation in the three body problem
\footnote{This section follows closely the corresponding review section given
in \cite{jim}}. The relativistic two-body bound state problem has a natural
origin from quantum field theory in the form of the Bethe-Salpeter equation
\cite{bse}. However, this equation is not usually applied in its full
four-dimensional form due to the difficulty of treating the relative time
coordinate \cite{nak}. Numerous three dimensional truncations of the
Bethe-Salpeter equation have been proposed for the relativistic two-body
problem \cite{quasi,yaes}. Some of these types of approximate methods have
previously been applied with considerable success to the $q\bar{q}$ meson
spectrum \cite{cra88}-\cite{crater2},\cite{jim} and \cite{rusger1}-\cite{yoon}.

The TBDE of constraint dynamics provide a manifestly covariant
three-dimensional truncation of the Bethe-Salpeter equation that more
efficiently distills two-body bound state and elastic scattering results.
\ Sazdjian \cite{saz} has shown that the Bethe-Salpeter equation can be
algebraically transformed into two independent equations. The first yields a
covariant three-dimensional eigenvalue equation which for spinless particles
takes the form
\begin{equation}
\biggl(\mathcal{H}_{10}+\mathcal{H}_{20}+2\Phi\biggr )\Psi(x_{1},x_{2})=0,
\label{eq:sum}%
\end{equation}
where $\mathcal{H}_{i0}=p_{i}^{2}+m_{i}^{2}$ . He finds that the
quasipotential\footnote{\ An earlier description of the connection of the
constraint approach to the quasipotential approach involving
Lippmann-Schwinger type of equations is given by Todorov \cite{tod} and Crater
et al \cite{becker} (see also \cite{rusger1}).} $\Phi$ is a modified geometric
series in the Bethe-Salpeter kernel $K$ such that in lowest order in $K$
\begin{equation}
\Phi=\pi iw\delta(P\cdot p)K, \label{bsen}%
\end{equation}
where $P=p_{1}+p_{2}$ is the total momentum, $p=\eta_{2}p_{1}-\eta_{1}p_{2} $
is the relative momentum, $w$ is the invariant total center of momentum (c.m.)
energy with $P^{2}=-w^{2}$. $\ $The $\eta_{i}$ must be chosen so that the
relative coordinate $x=x_{1}-x_{2}$ and $p$ are canonically conjugate, i.e.
$\eta_{1}+\eta_{2}=1$. The second independent equation overcomes the
difficulty of treating the relative time in the c.m. system by setting an
invariant condition on the relative momentum $p$,
\begin{equation}
(\mathcal{H}_{10}-\mathcal{H}_{20})\Psi(x_{1},x_{2})=0=2P\cdot p\Psi
(x_{1},x_{2}). \label{eq:dif}%
\end{equation}
Note that this implies $p^{\mu}\Psi=p_{\perp}^{\mu}\Psi\equiv(\eta^{\mu\nu
}+\hat{P}^{\mu}\hat{P}^{\nu})p_{\nu}\Psi$ in which $\hat{P}^{\mu}=P^{\mu}/w$
is a time like unit vector $(\hat{P}^{2}=-1)$ in the direction of the total momentum.

One can further combine the sum and the difference of Eqs. (\ref{eq:sum}) and
(\ref{eq:dif}) to obtain a set of two relativistic equations one for each
particle with each equation specifying two generalized mass-shell constraints
\begin{equation}
\mathcal{H}_{i}\Psi(x_{1},x_{2})=(p_{i}^{2}+m_{i}^{2}+\Phi)\Psi(x_{1}%
,x_{2})=0,~i=1,2, \label{dir}%
\end{equation}
including the interaction with the other particle by way of the quasipotential
$\Phi$. These constraint equations were originally derived using Dirac's
Hamiltonian constraint dynamics \cite{dirac,cnstr}. Dirac's constraint
dynamics stipulate that these two constraints must satisfy the compatibility
condition, $[\mathcal{H}_{1},\mathcal{H}_{2}]\Psi=0$, that is, they must be
first class \footnote{These constraint equations were originally proposed in
the form of classical generalized mass shell first class constraints
$\mathcal{H}_{i}=(p_{i}^{2}+m_{i}^{2}+\Phi_{i})\approx0$, and their
quantization $\mathcal{H}_{i}\Psi=0$ without reference to a quantum field
theory. For the classical $\mathcal{H}_{i}$ to be compatible, their Poisson
bracket with one \ another must either vanish strongly or depend on the
constraints themselves, $\{\mathcal{H}_{1},\mathcal{H}_{2}\}\approx0$. \ The
simplest solution of this equation is $\Phi_{1}=\Phi_{2}$, a kind of
relativistic third law condition, together with their common transverse
coordinate dependence $\Phi_{w}(x_{\perp}),$ just as with its quantum
version.}.\ With no external potentials, the coordinate dependence of the
quasipotential $\Phi$ $\ $would be through $x$ and the compatibility condition
becomes $[p_{1}^{2}-p_{2}^{2},\Phi]\Psi=P^{\mu}\partial\Phi/\partial x^{\mu
}=0$. In order for this to be true in general, $\Phi$ must depend on the
relative coordinate $x$ only through its component, $x_{\perp},$ perpendicular
to $P,$%
\begin{equation}
x_{\perp}^{\mu}=(\eta^{\mu\nu}+\hat{P}^{\mu}\hat{P}^{\nu})(x_{1}-x_{2})_{\nu}.
\label{ti}%
\end{equation}
Since the total momentum is conserved, the single component wave function
$\Psi~$in coordinate space is a product of a plane wave eigenstate of $P$ and
an internal part $\psi$ \cite{cra87}, depending on this $x_{\perp}%
.$\footnote{We use the same symbol $P$ for the eigenvalue so that the $w$
dependence in Eq. (\ref{em}) is regarded as an eigenvalue dependence. \ The
wave function $\Psi$ can be viewed as a relativistic 2-body wave function.}

We find a plausible structure for the two-body quasipotential $\Phi$ by
examining how scalar and vector interactions are introduced in the one-body
Klein-Gordon equation $(p^{2}+m^{2})\psi=(\mathbf{p}^{2}-\varepsilon^{2}%
+m^{2})\psi=0.$ This takes the form $(\mathbf{p}^{2}-\varepsilon^{2}%
+m^{2}+2mS+S^{2}+2\varepsilon A-A^{2})\psi=0~$when one introduces a scalar
interaction and timelike vector interaction via $m\rightarrow m+S~$and
$\varepsilon\rightarrow\varepsilon-A$. In the two-body case, separate
classical \cite{fw} and quantum field theory \cite{saz97} arguments show that
when one includes world scalar\ and vector interactions between the two
particles, then $\Phi$ depends on two underlying but unspecified invariant
functions $S(r)$ and $A(r)$ through the two-body Klein-Gordon-like potential
form with the same general structure, that is%
\begin{equation}
\Phi=2m_{w}S+S^{2}+2\varepsilon_{w}A-A^{2}. \label{em}%
\end{equation}
Those field theory based arguments point to the following c.m. energy
dependent forms
\begin{equation}
m_{w}=m_{1}m_{2}/w, \label{mw}%
\end{equation}
and%
\begin{equation}
\varepsilon_{w}=(w^{2}-m_{1}^{2}-m_{2}^{2})/2w. \label{ew}%
\end{equation}
They were first introduced by Todorov \cite{tod71} as the relativistic reduced
mass and effective particle energy for the two-body body system.\ Similar to
what happens in the nonrelativistic two-body problem, in the relativistic
case\ we have the motion of this effective particle taking place as if it were
in an external field \ (here generated by $S$ and $A$). \ The two kinematical
variables (\ref{mw}) and (\ref{ew}) are related to one another by the Einstein
condition
\begin{equation}
\varepsilon_{w}^{2}-m_{w}^{2}=b^{2}(w),
\end{equation}
where the invariant
\begin{equation}
b^{2}(w)\equiv(w^{4}-2w^{2}(m_{1}^{2}+m_{2}^{2})+(m_{1}^{2}-m_{2}^{2}%
)^{2})/4w^{2}, \label{bb}%
\end{equation}
is the c.m. value of the square of the relative momentum expressed as a
function of $w$. \ One also has%
\begin{equation}
b^{2}(w)=\varepsilon_{1}^{2}-m_{1}^{2}=\varepsilon_{2}^{2}-m_{2}^{2},
\end{equation}
in which $\varepsilon_{1}$ and $\varepsilon_{2}$ are the invariant c.m.
energies of the individual particles satisfying%
\begin{equation}
\ \varepsilon_{1}+\varepsilon_{2}=w,\ \varepsilon_{1}-\varepsilon_{2}%
=(m_{1}^{2}-m_{2}^{2})/w. \label{es}%
\end{equation}
In terms of these invariants, the relative momentum appearing in Eq.
(\ref{bsen}) and (\ref{eq:dif}) is given by%
\begin{equation}
p^{\mu}=(\varepsilon_{2}p_{1}^{\mu}-\varepsilon_{1}p_{2}^{\mu})/w\mathrm{,}
\label{relm}%
\end{equation}
so that $\eta_{1}+\eta_{2}=(\varepsilon_{1}+\varepsilon_{2})/w=1$. In
\cite{tod01} the forms for these two-body and effective particle variables are
given sound justifications based solely on relativistic kinematics,
supplementing the dynamical arguments of \cite{fw} and \cite{saz97}. In
summary, the wave function $\Psi(x_{1},x_{2})$ for spinless two body systems
satisfies \
\begin{align}
P\cdot p\Psi(x_{1},x_{2})  &  =0,\nonumber\\
(p^{2}+\Phi)\Psi(x_{1},x_{2})  &  =b^{2}\Psi(x_{1},x_{2}). \label{2s}%
\end{align}

Originally, the Two Body Dirac equations of constraint dynamics arose from a
supersymmetric treatment of two pseudoclassical constraints (with Grassmann
variables in place of gamma matrices) which were then quantized \cite{cra82}.
Sazdjian later derived \cite{saz} different forms of these same equations,
just as with their spinless counterparts above, as a covariant
three-dimensional truncation of the Bethe-Salpeter equation. \ The forms of
the equations are varied but the one that is the most familiar is the
"external potential" form similar in structure to the ordinary Dirac
equation\footnote{So-called hyperbolic forms of the Two Body Dirac equations
display more directly the connection between the source of the interactions
and the matrix structure of the three point vertex interactions of quantum
field theory. \ See \cite{jmath} and \cite{long}.}. \ For two particles
interacting through world scalar and vector interactions they are
\begin{align}
\mathcal{S}_{1}\psi &  \equiv\gamma_{51}(\gamma_{1}\cdot(p_{1}-\tilde{A}%
_{1})+m_{1}+\tilde{S}_{1})\Psi=0,\nonumber\\
\mathcal{S}_{2}\psi &  \equiv\gamma_{52}(\gamma_{2}\cdot(p_{2}-\tilde{A}%
_{2})+m_{2}+\tilde{S}_{2})\Psi=0. \label{tbde}%
\end{align}
Here $\Psi$ is a 16 component wave function consisting of an external plane
wave part that is an eigenstate of $P$ and an internal part $\psi
=\psi(x_{\perp})$. The vector potential$\ \tilde{A}_{i}^{\mu}$ is taken to be
an electromagnetic-like four-vector potential with the time and spacelike
portions both arising from a single invariant function $A$\footnote{In a
perturbative context, i.e. for weak potentials, that would mean that this
aspects of $\tilde{A}_{i}^{\mu}$ is regarded as arising from a Feynman gauge
vertex coupling of a form proportional to $\gamma_{1}^{\mu}\gamma_{2\mu}A$ .}.
\ The tilde on these four-vector potentials as well as on the scalar ones
$\tilde{S}_{i}$ indicate that these are not only position dependent but also
spin-dependent by way of the gamma matrices. \ The operators $\mathcal{S}_{1}$
and $\mathcal{S}_{2}$ must commute or at the very least $[\mathcal{S}%
_{1},\mathcal{S}_{2}]\psi=0$ since they operate on the same wave function
\footnote{The $\gamma_{5}$ matrices for each of the two particles are
designated by $\gamma_{5i}$ $i=1,2$. \ The reason for putting these matrices
out front of the whole expression is that including them facilitates the proof
of the compatibility condition, see \cite{cra82}.}. This compatibility
condition gives restrictions on the spin dependence\ which the vector and
scalar potentials%
\begin{equation}
\tilde{A}_{i}^{\mu}=\tilde{A}_{i}^{\mu}(A(r),p_{\perp},\hat{P},w,\gamma
_{1},\gamma_{2}),~\ \tilde{S}_{i}=\tilde{S}_{i}(S(r),A(r),p_{\perp},\hat
{P},w,\gamma_{1},\gamma_{2}). \label{paul1}%
\end{equation}
are allowed to have \footnote{The dependence of the scalar potentials
$\tilde{S}_{i}$ on the invariant $A(r)$ responsible for the
electromagnetic-like potential is seen in \cite{cra87} and \cite{saz97} to
result from the way the scalar and vector fields combine. That combination
leads to a two-body Klein-Gordon-like potential portion of $\Phi_{w}$ to be of
the form given in Eq. (\ref{em}).} in addition to requiring that they depend
on the invariant separation $r\equiv\sqrt{x_{\perp}^{2}}$ through the
invariants $A(r)$ and $S(r)$ . The covariant constraint (\ref{eq:dif}) can
also be shown to follow from Eq. (\ref{tbde}). \ We give the explicit
connections between $\tilde{A}_{i}^{\mu},\tilde{S}_{i}$ and the invariants
$A(r),$ and $S(r)$ in Appendix A of \cite{whitney}. The Pauli reduction of
these coupled Dirac equations lead to a covariant Schr\"{o}dinger-like
equation for the relative motion with an explicit spin-dependent potential
$\Phi,$
\begin{equation}
{\bigg(}p_{\perp}^{2}+\Phi(A(r),S(r),p_{\perp},\hat{P},w,\sigma_{1},\sigma
_{2}){\bigg)}\psi_{+}=b^{2}(w)\psi_{+}\ , \label{schlike}%
\end{equation}
with $b^{2}(w)$ playing the role of the eigenvalue.\footnote{Due to the
dependence of $\Phi_{w}$ on $w,$ this is a nonlinear eigenvalue equation. \ }
This eigenvalue equation can then be solved for the four-component effective
particle spinor wave function $\psi_{+}$ related to the 16 component spinor
$\psi(x_{\perp})$ (See Appendix A of \cite{jim}). \ In Ref. \cite{jim} a
number of important and desirable features of the set of Eq. (\ref{tbde}) and
the equivalent Schr\"{o}dinger-like equation (\ref{schlike}) are discussed. \ 

In \cite{crater2} we presented details of the application of this formalism to
meson spectroscopy using a covariant version of the Adler-Piran static quark
potential. Note especially that the equations used there displayed a
\textit{single} $\Phi(A(r),S(r),p_{\perp},\hat{P},w,\sigma_{1},\sigma_{2},)$
in Eq.\ (\ref{schlike}). It depends on the quark masses through factors such
as those that appear in Eq. (\ref{em}). However its dependence is the same for
all quark mass ratios - hence a single structure for all the $Q\bar{Q}%
,\ q\bar{Q},$ and $q\bar{q}$ mesons in a single overall fit. We found that the
fit provided by the TBDE for the entire meson spectrum (from the pion to the
excited bottomonium states) competes with the best fits to partial spectra
provided by other approaches and does so with the smallest number of
interaction functions (just $A(r)$ and $S(r)$) without additional cutoff
parameters necessary to make those approaches numerically tractable. We also
found that the pion bound state displays some characteristics of a Goldstone
boson. That is, as the quark mass tends to zero, the pion mass (unlike the
$\rho$ and the excited $\pi$) vanishes, in contrast to almost every other
relativistic potential model.

In Appendix A of \cite{jim} we outline the steps needed to obtain the explicit
c.m. form of \ Eq. (\ref{schlike}). \ That form is \cite{saz94}, \cite{liu},
\cite{crater2},
\begin{align}
&  \{\mathbf{p}^{2}+\Phi(\mathbf{r,}m_{1},m_{2},w,\mathbf{\sigma}%
_{1},\mathbf{\sigma}_{2})\}\psi_{+}=\nonumber\\
= &  \{\mathbf{p}^{2}+2m_{w}S+S^{2}+2\varepsilon_{w}A-A^{2}+\Phi
_{D}\nonumber\\
&  +\mathbf{L\cdot(\sigma}_{1}\mathbf{+\sigma}_{2}\mathbf{)}\Phi
_{SO}+\mathbf{\sigma}_{1}\mathbf{\cdot\hat{r}\sigma}_{2}\mathbf{\cdot\hat
{r}L\cdot(\sigma}_{1}\mathbf{+\sigma}_{2}\mathbf{)}\Phi_{SOT}\nonumber\\
&  +\mathbf{\sigma}_{1}\mathbf{\cdot\sigma}_{2}\Phi_{SS}+(3\mathbf{\sigma}%
_{1}\mathbf{\cdot\hat{r}\sigma}_{2}\mathbf{\ \cdot\hat{r}-\sigma}%
_{1}\mathbf{\cdot\sigma}_{2})\Phi_{T}\nonumber\\
&  +\mathbf{L\cdot(\sigma}_{1}\mathbf{-\sigma}_{2}\mathbf{)}\Phi
_{SOD}+i\mathbf{L\cdot\sigma}_{1}\mathbf{\times\sigma}_{2}\Phi_{SOX}\}\psi
_{+}\nonumber\\
&  =b^{2}\psi_{+}.\label{57}%
\end{align}
Thus is derived a relativistic two-body Schr\"{o}dinger-like equation for
world scalar and vector interactions. \ The minimal $2m_{w}S+S^{2}%
+2\varepsilon_{w}A-A^{2}$ portion is the classical interaction potential part
(which also appears in the spinless Klein-Gordon equations), the
$\mathbf{L\cdot(\sigma}_{1}\mathbf{\pm\sigma}_{2}\mathbf{)}$ terms represent
magnetic dipole moment interactions with an effective magnetic field and
Thomas precession, and $\mathbf{\sigma}_{1}\mathbf{\cdot\hat{r}\sigma}%
_{2}\mathbf{\cdot\hat{r}},$\textbf{\ }$\mathbf{\sigma}_{1}\mathbf{\cdot\sigma
}_{2}\ ~$terms arise from dipole-dipole interactions and their relativistic
corrections. \ A main focus of this work will be to derive a similar equation
to Eq. (\ref{57}) for the three-body baryon system as a whole. The detailed
\ forms of the separate quasipotentials $\Phi$ are given in Appendix A of
\cite{jim}. \ The subscripts of most of the quasipotentials are self
explanatory \footnote{The subscript on quasipotential $\Phi_{D}$ refers to
Darwin. \ It consist of what are called Darwin terms, those that are the
two-body analogue of terms that accompany the spin-orbit term in the one-body
Pauli reduction of the ordinary one-body Dirac equation, and ones related by
canonical transformations to Darwin interactions \cite{fw,sch73}, momentum
dependent terms arising from retardation effects. The subscripts on the other
quasipotentials refer respectively to $SO$ (spin-orbit), $SOD$ (spin-orbit
difference), $SOX$ (spin-orbit cross terms), $SS$ (spin-spin), $T$ (tensor),
$SOT$ (spin-orbit-tensor)}. \ After the eigenvalue $b^{2}$ of (\ref{57}) is
obtained, the invariant mass of the composite two-body system $w$ can then be
obtained by inverting Eq.\ (\ref{bb}). It is given explicitly by
\begin{equation}
w=\sqrt{b^{2}+m_{1}^{2}}+\sqrt{b^{2}+m_{2}^{2}}.
\end{equation}
The structure of the linear and quadratic terms in Eq. (\ref{57}) as well as
the Darwin and spin-orbit terms, are plausible in light of the discussion
given above Eq. (\ref{em}), and in light of the static limit Dirac structures
that come about from the Pauli reduction of the Dirac equation. Their
appearance as well as that of the remaining spin structures are direct
outcomes of the Pauli reductions of the simultaneous TBDE Eq. (\ref{tbde}).

This is the framework for the two-body system in a fully relativistic
formalism, so from here we go to larger systems. \ Sazdjian\cite{5} has done
considerable work on the $N$-body system, which will be reviewed shortly.
\ Although he does not deal with spin dependence with as\ much detail as done
here, he does provide a very useful framework for the $N$-body problem in a
constraint formalism.

\subsection{Two Body Dirac Equations: Explicit Forms of the Potentials}

Since the forms of the potentials in the three-body case are similar to those
in the two-body case, it is of use to briefly describe the two-body
interacting potentials and how they affect the wavefunction. \ This section
then contains a review of how the operators of the tensor, spin-spin,
spin-orbit, spin-orbit difference and spin-orbit exchange work on a $\langle
jlsn|$ state coupling, where $n$ is the radial quantum number. \ In the
three-body case there will be two $n^{\prime}s$, one for each relative
coordinate, but here there is just one. \ 

First, our quasipotential (energy dependent effective potential) is defined by%

\begin{align}
\Phi &  =\Phi_{SI}+\Phi_{D}+\mathbf{L\cdot(}\boldsymbol{\sigma}_{1}%
\mathbf{+}\boldsymbol{\sigma}_{2}\mathbf{)}\Phi_{SO}+\boldsymbol{\sigma}%
_{1}\mathbf{\cdot\hat{r}}\boldsymbol{\sigma}_{2}\mathbf{\cdot\hat{r}L\cdot
(}\boldsymbol{\sigma}_{1}\mathbf{+}\boldsymbol{\sigma}_{2}\mathbf{)}\Phi
_{SOT}\label{333}\\
&  +\boldsymbol{\sigma}_{1}\mathbf{\cdot}\boldsymbol{\sigma}_{2}\Phi
_{SS}+(3\boldsymbol{\sigma}_{1}\mathbf{\cdot\hat{r}}\boldsymbol{\sigma}%
_{2}\mathbf{\ \cdot\hat{r}-}\boldsymbol{\sigma}_{1}\mathbf{\cdot
}\boldsymbol{\sigma}_{2})\Phi_{T}+\mathbf{L\cdot(}\boldsymbol{\sigma}%
_{1}\mathbf{-}\boldsymbol{\sigma}_{2}\mathbf{)}\Phi_{SOD}+i\mathbf{L\cdot
}\boldsymbol{\sigma}_{1}\mathbf{\times}\boldsymbol{\sigma}_{2}\Phi
_{SOX},\nonumber\\
\Phi_{SI}  &  =2m_{w}S+S^{2}+2\varepsilon_{w}A-A^{2}\nonumber
\end{align}
where the potential terms $\Phi_{D},\Phi_{SO},\Phi_{SOT},\Phi_{SS},\Phi
_{T},\Phi_{SOD},\Phi_{SOX}$ described earlier are all collections of two-body
terms depending on the masses, distances between the two particles, invariant
c.m. energies of the two particles and the invariant energy of the total
two-body system system. The explicit forms of these are therefore not
important to the current discussion of the operators and so will be left in
this form for simplicity's sake. \ The spin independent and Darwin terms have
no spin operators and so when used on a $\langle jlsn|$ state they just give%
\begin{align}
\langle jlsn|\Phi_{SI}|jl^{\prime}s^{\prime}n^{\prime}\rangle &
=\delta_{ll^{\prime}}\delta_{ss^{\prime}}\langle n|\Phi_{SI}|n^{\prime}%
\rangle,\nonumber\\
\langle jlsn|\Phi_{D}|jl^{\prime}s^{\prime}n^{\prime}\rangle &  =\delta
_{ll^{\prime}}\delta_{ss^{\prime}}\langle n|\Phi_{SI}|n^{\prime}\rangle.
\end{align}
The spin-orbit gives%
\begin{equation}
\langle jlsn|\mathbf{L\cdot(}\boldsymbol{\sigma}_{1}\mathbf{+}%
\boldsymbol{\sigma}_{2}\mathbf{)}\Phi_{SO}|jl^{\prime}s^{\prime}n^{\prime
}\rangle=[j(j+1)-l(l+1)-2]\delta_{ll^{\prime}}\delta_{ss^{\prime}}\delta
_{s1}\langle n|\Phi_{SO}|n^{\prime}\rangle.
\end{equation}
As we will show later, while in the two-body case this $l$,$j,$and $s$ are for
the entire system, in the three-body problem it is just for each pair of
particles and so this spin-orbit function requires additional (and extensive)
manipulation in order to reach a completely coupled $|JLS\rangle$ state
\textit{for each set of particles}. \ The emphasis here is important as this
is the main difficulty in going from the two-body formalism to a three-body
one, in this work as well as others.

The tensor and spin-orbit tensor terms allow for coupling of different $l$
states as well as identical $l$ states, as shown in(\cite{becker}), while the
spin-orbit difference and spin-orbit exchange \textit{only} allow couplings
between different spin\textit{\ }states. \ The exact derivations of these
potential terms are done in (\cite{jim,wongyoon,alstine2009}). \ Since one of
the goals of this work is to compare essentially the same methods that worked
well for the meson spectrum to the baryon spectrum, we use these same
potential terms in mostly the same form as they appear in the pure two-body
case. \ The two-body operators are therefore defined and described in
preparation for their adaptation to the three-body potential. \ Now we will
give definitions for the scalar and vector potentials used in our model and
from that define the two-body potentials $\Phi_{D},\Phi_{SO},\Phi_{SOT}%
,\Phi_{SS},\Phi_{T},\mathbf{\Phi}_{SOD},$ and $\Phi_{SOX}.$

\bigskip

\subsection{Explicit Forms of the QCD Model Potentials}

\qquad The authors of \cite{crater2} have used a sophisticated form of the
static quark potential developed by Adler and Piran \cite{adl}, one that has
ties at all length scales to field theoretic data and from this obtained good
agreement with the quarkonium spectrum from experimental data. \ However, it
is much more common in nonrelativistic treatments to use the static quark
Cornell potential\cite{crnl} for potential model studies,
\begin{equation}
V(r)=-\frac{\alpha_{c}}{r}+br,
\end{equation}
as in \cite{Bar92,Won01}. Although not displaying asymptotic freedom, it does
give the dominant Coulomb-like behavior as well as the linear quark
confinement. Early on a model was proposed by Richardson for a static
potential which both depends only a single scale size $\Lambda$ and
interpolates in a simple way between asymptotic freedom and linear confinement
\cite{rch}. Richardson's model for the static interquark potential in momentum
space is
\begin{equation}
\tilde{V}(\mathbf{q)=}-\frac{16\pi}{27}\frac{1}{\mathbf{q}^{2}\ln
(1+\mathbf{\ q}^{2}/\Lambda^{2})}, \label{rcr}%
\end{equation}
arising from the assumption that \
\begin{equation}
\tilde{V}(\mathbf{q)=}-\frac{4\alpha_{s}(\mathbf{q}^{2})}{3\mathbf{q}^{2}%
}\mathbf{,}%
\end{equation}
(including the color factor $-4/3$). \ It is important to note that this is
for a $q\overline{q}$ color singlet state for the meson spectrum. \ \ In order
to properly account for asymptotic freedom, we must have $\mathbf{q}%
^{2}/\Lambda^{2}>>1,$ which gives
\begin{equation}
\alpha_{s}(\mathbf{q}^{2})\rightarrow\frac{8\pi}{27}\frac{1}{\ln
(\mathbf{q}^{2}/\Lambda^{2})}.
\end{equation}
On the other hand, the property of linear confinement requires that for
$\Lambda r>>1,$ $V(r)\propto r$ or equivalently that for $\mathbf{q}%
^{2}/\Lambda^{2}<<1$ one must impose $\alpha_{s}\mathbf{(\mathbf{q}^{2})\sim
q}^{-2}$.\ The interpolation of Eq.\ (\ref{rcr}) is not tied at all in the
intermediate region and only roughly tied in the large $r$ region to any field
theoretic data. Nevertheless it provides a convenient one-parameter form for
the static quark potential. \ In coordinate space it has the form
\begin{equation}
V(r)=\frac{8\pi\Lambda^{2}r}{27}-\frac{8\pi f(\Lambda r)}{27r},
\end{equation}
where $f(\Lambda r)$ is given by a complicated integral transform\footnote{In
addition to the spin independent nonrelativistic model presented in \cite{rch}
see also a relativistic extension of it given in \cite{hva}.} that displays
the asymptotic freedom behavior for $r\rightarrow0$ of
\begin{equation}
f(\Lambda r)\rightarrow-\frac{1}{\ln\Lambda r},
\end{equation}
while for $r\rightarrow\infty,$
\begin{equation}
f(\Lambda r)\rightarrow1.
\end{equation}
\ 

A simpler model for the potential function $f(r)$, which we use in this paper
and one which displays the same large and small $r$ behavior is\footnote{An
earlier coordinate space form that displays asymptotic freedom as well as
linear quark confinement proposed in \cite{licht} is $V=(8\pi/27)(1-\lambda
r)^{2}/(r\ln\lambda r).$} \
\begin{equation}
V(r)=\frac{8\pi\Lambda^{2}r}{27}-\frac{16\pi}{27r\ln(e^{2}+1/(\Lambda r)^{2}%
)}. \label{rrich}%
\end{equation}
It amounts to replacing Richardson's $f(\Lambda r)$ by $2/\ln(e^{2}+1/(\Lambda
r)^{2}),$ having the same limits $\left(  e=\exp(1)\right)  $. \ The slightly
modified forms of the scalar and vector invariant potentials, including the
electromagnetic part
\begin{align}
S  &  =\frac{8\pi\Lambda^{2}r}{27}\label{DefinedPotentials}\\
A  &  =-\frac{16\pi}{27r\log(Ke^{2}+\frac{B}{(\Lambda r)^{2}})}+\frac
{e_{1}e_{2}}{4\pi r}\nonumber
\end{align}
\ are used to construct all of the individual $\Phi$ terms. \ Their explicit
forms, derived from the above $A$ and $S$ are given in Appendix A of
\cite{whitney}. In the case of the baryons these are slightly changed due to a
different color factor\cite{greenbook} ($-4\alpha_{s}/3$ becomes -$2\alpha
_{s}/3$ due to this being quark-quark and not quark-antiquark $\ $interactions
as with mesons$)$ to%
\begin{align}
S  &  =\frac{4\pi\Lambda^{2}r}{27}\nonumber\\
A  &  =-\frac{8\pi}{27r\log(Ke^{2}+\frac{B}{(\Lambda r)^{2}})}+\frac
{e_{1}e_{2}}{4\pi r} \label{sa}%
\end{align}
and also of course there is no longer just one interaction but three, so $r$
becomes $r_{12},r_{13},$or $r_{23}$, depending on which potential we are
currently discussing.\ \ The scalar confining interaction, \ unlike the vector
one, is \ not regarded as coming from fundamental vertices or potentials
involving current quarks. If it were treated as fundamental then the scalar
interaction would be repulsive within baryons if it is\ attractive within
mesons. \ Instead we treat \ the confining interaction as arising from
effective potentials between constituent quarks and use this freedom to allow
us, for phenomenological reasons, to choose the sign of the $qq$ scalar
potential to be the same sign as the $q\bar{q}$ scalar
potential\footnote{\ The problem with vector confining interactions is that 1)
they will generate long distance spin-spin interactions and 2) in the context
of Dirac like equations produce anticonfining ($-A^{2}$) terms. \ See
\cite{greenbook} for a more detailed \ discussion of this problem. \ See also
\cite{swanson} where it is shown that the Dirac structure of confinement for
mesons could be of a timelike-vector nature in the heavy quark limit of QCD.
\ This would alleviate the problem of 1). \ Further, they find that
nonperturbative mixing between ordinary and hybrid $Q\bar{Q}$ states seems to
allow spin orbit effects as if arising from confining scalar interactions.
\ \ }. \ 

The technique that Crater et al. used in the two body problem \cite{becker}
for finding the eigenvalues is called the Inverse Power Method and its
application depends on the variables being separable. \ Unlike the two-body
problem, the variables are not separable in the three-body problem. \ This
requires the use of the variational principle, which in turn requires a basis
which we will describe in a later section.\ We turn now to a discussion of the
three body problem.

\section{The Three Body Problem}

\qquad Now that we have completed our review of constraint dynamics and
associated potentials for the relativistic two-body problem, we move on to the
relativistic three-body one. \ The approach to the $N$-body problem that we
use are those developed H.\ Sazdjian \cite{5}. They are not directly solvable
for more than $N=2$, except for confined systems in which the problems of
cluster decomposition do not need to be addressed. \ In this paper we adapt
our two-body constraint formalism to his formalism for $N=3$ and from there we
obtain to a Schr\"{o}dinger-like form for the three body system, as we did in
the two-body problem.\ 

\subsection{Sazdjian's N body formalism and the Three Body Problem}

\qquad This section will focus on reviewing Sazdjian's work on the two-body
and $N$-body systems\cite{5}. \ We describe his derivations for the $N$-body
problem and distill them down to a three-body formalism that we can then use
for bound states of quarks in baryons. \ Sazdjian begins by applying the
covariant formalism with $N$ constraints for the $N$-particle case of the form%
\begin{equation}
\mathcal{H}_{a}\psi=(p_{a}^{2}+m_{a}^{2}+\Phi_{a})\psi=0, \label{83}%
\end{equation}
The compatibility condition is then%
\begin{equation}
\lbrack\mathcal{H}_{a},\mathcal{H}_{b}]\Psi=0\text{ \ \ \ (}a,b=1,...,N),
\label{gc}%
\end{equation}
which are $N(N-1)/2$ in number and give conditions on the interaction
potentials ($\Phi_{a})$. \ However, these equations, unlike the two-body ones
\ (where $\Phi$ is a function of $x_{\bot}$) have no closed \ form solutions.
Furthermore, the two body potentials would become non-local operators, due to
the two-body momentum operators $P_{ab}=p_{a}+p_{b}$ no longer representing
the total momentum of the system; since they would not be constants of the
motion they would not possess corresponding eigenvalues. \ On the other hand
the total momentum in two-body case is a number (an eigenvalue). \ Sazdjian
abandons this approach and instead takes a simpler one that works only for
confined systems where questions of correct cluster decompositions do not have
to be addressed.

Sazdjian begins by working with the free $N-$body system where one can be
guided by the simplifying features of the two-body system. \ Since the system
must reduce to that of the free case in the absence of interactions, he found
it useful to begin with the $N-$body system without any interacting
potentials. A review of the details of his approach is given in Appendix B of
\cite{whitney}, but the end result is a single $N-$body wave equation for the
system as a whole and set of $N$ equations for the individual invariant c.m.
particle energies $\varepsilon_{a},$ \ This latter equation is%
\begin{equation}
N\varepsilon_{a}-\sum_{b}\frac{(m_{a}^{2}-m_{b}^{2})}{(\varepsilon
_{a}+\varepsilon_{b})}=w,(a=1,...,N), \label{85}%
\end{equation}
where $w$ is the total invariant c.m. energy%
\begin{equation}
w=\sum_{b}\varepsilon_{b}.
\end{equation}
These equations cannot be solved exactly for the $\varepsilon_{b}$ simply
except in the two-body case, which reduce down to Eq.(\ref{es})%
\begin{equation}
\varepsilon_{1}-\varepsilon_{2}=\frac{m_{1}^{2}-m_{2}^{2}}{w},
\end{equation}
as expected. \ He shows it is possible, however, to find an approximate
solution \ for the $\varepsilon_{b}$ by using successive iterations in the
general case given by%
\begin{equation}
\varepsilon_{a}=\frac{w}{N}+\frac{1}{N}\sum_{b\neq1}^{N}\frac{(m_{a}-m_{b}%
)}{\left[  1+(w-M)/2m_{a}m_{b}\sum_{c=1}^{N}1/2m_{c}\right]  };~~a=1,..N.
\label{en}%
\end{equation}
The full $N-$body equation which utilizes these c.m. energy eigenvalues is%
\begin{equation}
\sum_{a=1}^{N}[-\varepsilon_{a}^{2}+N\frac{p_{a\perp}^{2}/(2\varepsilon_{a}%
)}{\sum_{b=1}^{N}1/2\varepsilon_{b}}+m_{a}^{2}]\psi=0. \label{88}%
\end{equation}
This equation (\ref{88}) determines the total c.m. energy $w$ in terms of the
masses and the transverse momenta. \ For $N=2$, \ this becomes the free form
of the two-body system earlier given in Eq. (\ref{2s})
\begin{align}
(p_{\perp}^{2}-b^{2})\psi &  =0.\nonumber\\
b^{2}  &  =\varepsilon_{1}^{2}-m_{1}^{2}=\varepsilon_{2}^{2}-m_{2}^{2}%
\end{align}
Finally, he also gives the system of $N$ Klein-Gordon equations\qquad%
\begin{equation}
\{-\varepsilon_{a}^{2}+[\sum_{b=1}^{N}\frac{p_{b\perp}^{^{2}}}{(2\varepsilon
_{b})}]/(\sum_{c=1}\frac{1}{2\varepsilon_{c}})+m_{a}^{2}\}\Psi=0\text{
\ \ \ \ \ (}a=1,...,N). \label{89}%
\end{equation}

Sazdjian finds that in the interacting case, the structure of these $N$
non-independent wave equations as well as Eq. (\ref{88}) are kinematic in
nature and should not be modified by the interactions. For example, particle
$a$ "feels" an interaction potential $\Phi_{a}$ \ that enters additively into
its kinetic energy term by the relation%
\begin{equation}
p_{\perp a}^{2}\rightarrow p_{\perp a}^{2}+\Phi_{a}.
\end{equation}
Eq.(\ref{88}) then becomes
\begin{equation}
\sum_{a}[-\varepsilon_{a}^{2}+N\frac{(p_{a\perp}^{2}+\Phi_{a})/(2\varepsilon
_{a})}{\sum_{b}1/2\varepsilon_{b}}+m_{a}^{2}]\psi=0, \label{302}%
\end{equation}
which in the two-body case is
\begin{equation}
2[\frac{p_{1\perp}^{2}+\Phi_{1}}{w}\varepsilon_{2}]+2[\frac{p_{2\perp}%
^{2}+\Phi_{2}}{w}\varepsilon_{1}]\psi=\left(  \varepsilon_{1}^{2}-m_{1}%
^{2}+\varepsilon_{2}^{2}-m_{2}^{2}\right)  \psi.
\end{equation}
Since $p_{1\perp}^{2}=p_{2\perp}^{2}$ and $\Phi_{1}=\Phi_{2}$ this reduces to
Eq. (\ref{2s}). \ In the general $N$-body case, the individual wave equations
(\ref{89}) become%
\begin{equation}
\{-\varepsilon_{a}^{2}+[\sum_{b=1}^{N}\frac{(p_{b\perp}^{2}+\Phi_{b}%
)}{(2\varepsilon_{b})}]/(\sum_{c}\frac{1}{2\varepsilon_{c}})+m_{a}^{2}%
\}\Psi=0\text{ \ \ \ \ \ (}a=1,...,N). \label{301}%
\end{equation}
A sufficient condition for compatibility of these wave equations is to take%
\begin{align}
\Phi_{a}  &  =\sum_{b\neq a}^{N}\Phi_{ab}(x_{ab\perp}),\nonumber\\
x_{ab\perp}^{\mu}  &  =(x_{a}^{\mu}-x_{b}^{\mu})-P^{\mu}\hat{P}\cdot
(x_{a}^{\mu}-x_{b}^{\mu}), \label{3p}%
\end{align}
where again, $P$ is the \textit{total }momentum\footnote{Note that this
dependence of the potential on the part of the potential that depends on
component of the relative coordinates perpendicular to the momentum of the
total system, is allowed as long as the issue of cluster decomposition need
not be addressed. \ In that event, where one could separate out a part of the
system from the remaining part, it is not meaningful to require the potentials
to depend on the total momemtum of the total system instead of that of its
pairs of subconstituents.}%
\begin{equation}
P=\sum_{a=1}^{N}p_{a}.
\end{equation}

We choose the $\Phi_{ab}$ in this equation (Eq. (\ref{3p})) to have the same
functional dependence on $S$ and $A$ as in Eq. (\ref{333}) our two-body Dirac
approach. \ Sazdjian does not deal directly with the the explicit form of the
potential. \ Our choice in this work is to use the potential from the TBDE in
the Sazdjian three-body equations. \ In this paper we do not include three
body forces.

\subsubsection{Our Adaptation of Sazdjian's Three-Body Generalization}

\qquad In order to apply the work done by\ Sazdjian to our problem, we have to
specialize his $N$-body equations to the three-body system and derive the
appropriate effective Hamiltonian, eventually ending up with an equation that
looks very much like a nonrelativistic three body Schr\"{o}dinger equation,
reducing to it in the nonrelativistic limit. \ We obtain the following
approximation for the three-body eigenvalue equation, which comes from
specializing the expanded three body version of Eq.(\ref{302}) as follows:
\begin{align}
0 &  =[\varepsilon_{1}^{2}-m_{1}^{2}-3\frac{p_{1\perp}^{2}+\Phi_{1}%
}{\varepsilon_{1}(1/\varepsilon_{1}+1/\varepsilon_{2}+1/\varepsilon_{3}%
)}\nonumber\\
&  +\varepsilon_{2}^{2}-m_{2}^{2}-3\frac{p_{2\perp}^{2}+\Phi_{2}}%
{\varepsilon_{2}(1/\varepsilon_{1}+1/\varepsilon_{2}+1/\varepsilon_{3}%
)}\nonumber\\
&  +\varepsilon_{3}^{2}-m_{3}^{2}-3\frac{p_{3\perp}^{2}+\Phi_{3}}%
{\varepsilon_{3}(1/\varepsilon_{1}+1/\varepsilon_{2}+1/\varepsilon_{3})}%
]\psi(x_{12\perp},x_{23\perp},x_{31\perp}),\label{1st3bodyy}%
\end{align}
in which the epsilons, representing the c.m. energy of each quark, are given
to a good approximation by Eq. (\ref{en}). The potentials $\Phi_{i}$ are
linear combinations of the two-body interacting potentials
\begin{align}
\Phi_{1} &  =\Phi_{12}(x_{12\perp},\varepsilon_{1},\varepsilon_{2})+\Phi
_{23}(x_{23\perp},\varepsilon_{2},\varepsilon_{3}),\nonumber\\
\Phi_{2} &  =\Phi_{23}(x_{23\perp},\varepsilon_{2},\varepsilon_{3})+\Phi
_{31}(x_{31\perp},\varepsilon_{3},\varepsilon_{1}),\nonumber\\
\Phi_{3} &  =\Phi_{31}(x_{31\perp},\varepsilon_{3},\varepsilon_{1})+\Phi
_{12}(x_{12\perp},\varepsilon_{1},\varepsilon_{2}).
\end{align}
This equation (\ref{1st3bodyy}) is essentially the three-body version of the
two-body equation:
\begin{equation}
\mathcal{H}=\frac{(p_{1}^{2}+m_{1}^{2}+\Phi)}{2\varepsilon_{1}}+\frac
{(p_{2}^{2}+m_{2}^{2}+\Phi)}{2\varepsilon_{2}},
\end{equation}
as long as one restricts oneself to confining interactions.

In the above equation (\ref{1st3bodyy})
\begin{align}
p_{i\perp}  &  =p_{i}+p_{i}\cdot\hat{P}\hat{P},\nonumber\\
x_{ij\perp}  &  =x_{ij}+x_{ij}\cdot\hat{P}\hat{P},\nonumber\\
\hat{P}  &  =\frac{P}{\sqrt{-P^{2}}},\nonumber\\
P  &  =p_{1}+p_{2}+p_{3}.
\end{align}
We define%
\begin{equation}
E=w-M\equiv\varepsilon_{1}+\varepsilon_{2}+\varepsilon_{3}-m_{1}-m_{2}-m_{3},
\end{equation}
in order to bring (\ref{1st3bodyy}) into a more usable and familiar
(Schr\"{o}dinger-like) form. \ This form is given by \
\begin{equation}
\mathcal{H\psi}\equiv\frac{1}{F}\left(  \frac{p_{1\perp}^{2}+\Phi_{12}%
+\Phi_{13}}{2\varepsilon_{1}(E,m_{1},m_{2},m_{3})}+\frac{p_{2\perp}^{2}%
+\Phi_{23}+\Phi_{12}}{2\varepsilon_{2}(E,m_{1},m_{2},m_{3})}+\frac{p_{3\perp
}^{2}+\Phi_{31}+\Phi_{23}}{2\varepsilon_{3}(E,m_{1},m_{2},m_{3})}\right)
\psi=E\psi, \label{duranham}%
\end{equation}
where the function $F=F(w,m_{1},m_{2},m_{3}),$ an invariant function of the
total energy of the system and the masses of the particles\cite{Duran}, is the
result of an algebraic manipulation (details and explicit form given in
Appendix C of \cite{whitney} ) . \ The functional forms of the $\varepsilon
_{i}$ are given in Eq. (\ref{en}). \ The effects of spin are included by
choosing for the $\Phi_{ij}$ given in Eq. (\ref{333}). Eq. (\ref{duranham})
serves as our basic three-body bound state equation for quarks in the baryon.
\ Even though the structure of the equation is nonrelativistic it is Lorentz
invariant as it is composed of invariant portions. \ They are of two sorts,
the square of space-like and time-like vectors represented respectively by
$p_{i\perp}^{2}$ and $x_{ij\perp}^{2}$\ one the one hand and $P^{2}=-w^{2}$ on
the other. \ In the c.m. frame the former become the squares $\mathbf{p}%
_{i}^{2}$ and $\mathbf{x}_{ij}^{2}$ of three vectors. \ In the
non-relativistic limit when $\left\vert E\right\vert <<m_{i}$, then the
invariant $F\rightarrow1,$ $\varepsilon_{i}\rightarrow m_{i}$ and the operator
$\mathcal{H}$ in Eq. (\ref{duranham}) becomes an ordinary non-relativistic
Hamiltonian. \ 

So now, we have gone from\ Sazdjian's N-body formalism to a condensed
three-body one that is easy to work with in the constraint dynamics approach,
as the Hamiltonian is now in a familiar form. \ This equation has the distinct
advantage that it is like (for the purposes of solving it anyway) a
non-relativistic Schr\"{o}dinger equation. \ It is, of course, still
relativistic, but it is now in a form that is much more easily recognizable
and usable than Eq.(\ref{301}). \ It is important to note the recursive nature
of this equation as this becomes highly relevant in the numerical studies.
\ The $\Phi$'s are dependent on the $\varepsilon^{\prime}s$ and $w$ and so we
must begin with an initial guess and solve the equation iteratively until an
acceptable level of convergence of $w$ is met.

\section{The relativistic three body problem for baryons}

\qquad It should be noted that models such as our two body and its three body
relativistic generalizations are often referred to as "naive quark models" in
that they do not account for the swarm of gluons and sea quark-anti-quark pair
interactions directly. \ Rather, the model has all of the interacting forces
existing only between each quark-quark pair. \ \ 

The process of going from a two-body system to a three-body one is not as
straightforward as one might expect. \ Since the interacting potentials are
limited to each quark-quark pair, there are now three times as many terms.
\ Additionally, there are three sets of coordinates ($\mathbf{r}%
_{1}-\mathbf{r}_{2},\mathbf{r}_{2}-\mathbf{r}_{3},\mathbf{r}_{1}%
-\mathbf{r}_{3})$ instead of just one distance between the quarks as in
the\ two-body case. \ This is best treated with a relative coordinate
substitution that reduces the number of relative coordinates from three to
two. \ The coordinate transform used in this work is similar to and uses the
same notation as Capstick and Isgur's (\cite{2}) work, but is not an identical
transformation due to treating the more general case of all three quarks as
possibly having different masses, in particular not choosing the $u$ and $d$
quarks to be identical in mass. \ It also uses invariant c.m. energies
$\varepsilon_{a}$ in place of masses.

The sections that follow describe the methods used in going from a two-body
system to a three-body one, mostly dealing with the potentials and coordinate
transforms. \ We also describe our Gaussian basis functions and the reasoning
behind them. \ Referring to Eq.(\ref{333})), the $\Phi_{SI}$ and Darwin
($\Phi_{D})$ terms now expand simply from one term to three (to account for
all three two-body interactions) and their matrix elements are no more
complicated in principle than what occurs in the two body problem. \ However,
matrix elements for the the spin-spin ($\Phi_{SS})$, spin-orbit ($\Phi
_{SO},\Phi_{SOT},\Phi_{SOX})$ and tensor ($\Phi_{T})$ terms require
manipulations related to total $J$, total $L$ and total $S$ and are
considerably more complex than what appears in the that the two-body system. \ 

\subsection{Spin-Flavor-Space States}

\qquad Here we list all of the spin-flavor states for all the baryons in our
fit. \ They are composed of products of spin wavefunctions, denoted $\chi$,
and flavor wavefunctions, denoted as $\phi$ \cite{greenbook}. The flavor wave
functions are not listed for charmed or bottom baryons as those are the same
wave functions with a $b$ or $c$ quark in place of a $u,d,$ or $s,$ depending
on the baryon. \ The spin wavefunctions are, explicitly \
\begin{align}
\chi^{s}(S_{z}  &  =\frac{3}{2})=\uparrow\uparrow\uparrow,\nonumber\\
\chi^{\prime}(S_{z}  &  =\frac{1}{2})=\frac{1}{\sqrt{2}}(\uparrow
\downarrow\uparrow-\downarrow\uparrow\uparrow),\nonumber\\
\chi^{\prime\prime}(S_{z}  &  =\frac{1}{2})=\frac{1}{\sqrt{6}}(2\uparrow
\uparrow\downarrow-\uparrow\downarrow\uparrow-\downarrow\uparrow\uparrow).
\end{align}
\ There are four different flavor wavefunctions, denoted as $\phi^{\prime
},\phi^{\prime\prime},\phi^{s},\phi^{a},$ corresponding here to the ground
state octet and decimet baryons and their extensions to include charm and
bottom quarks. \ Note that $\phi^{s}$ is a symmetric linear combination of the
listed quarks (e.g. $uud=\frac{1}{\sqrt{3}}[uud+udu+duu])$%

\begin{table}[tbp] \centering
\caption{Baryon flavor wavefunctions}$%
\begin{tabular}
[c]{|l|l|l|l|}\hline
& $\phi^{s}$ & $\phi^{^{\prime}}$ & $\phi^{\prime\prime}$\\\hline
p &  & $\frac{1}{\sqrt{2}}(udu-duu)$ & $\frac{1}{\sqrt{6}}(2uud-duu-udu)$%
\\\hline
n &  & $\frac{1}{\sqrt{2}}(udd-dud)$ & $\frac{1}{\sqrt{6}}(dud-udd-2ddu)$%
\\\hline
$\Lambda$ &  & $\frac{1}{2\sqrt{3}}(usd+sdu-sud-dsu-2dus+2uds)$ & $\frac{1}%
{2}(sud+usd-sdu-dsu)$\\\hline
$\Delta^{++}$ & $uuu$ &  & \\\hline
$\Delta^{+}$ & $uud$ &  & \\\hline
$\Delta^{0}$ & $udd$ &  & \\\hline
$\Delta^{-}$ & $ddd$ &  & \\\hline
$\Sigma^{+}$ & $uus$ & $\frac{1}{\sqrt{2}}(suu-usu)$ & $\frac{1}{\sqrt{6}%
}(suu-usu-2uus)$\\\hline
$\Sigma^{0}$ & $uds$ & $\frac{1}{2}(sud+sdu-usd-dsu)$ & $\frac{1}{2\sqrt{3}%
}(usd+sdu+sud+dsu-2dus-2uds)$\\\hline
$\Sigma^{-}$ & $dds$ & $\frac{1}{\sqrt{2}}(sdd-dsd)$ & $\frac{1}{\sqrt{6}%
}(sdd-dsd-2dds)$\\\hline
$\Xi^{0}$ & $uss$ & $\frac{1}{\sqrt{2}}(sus-uss)$ & $\frac{1}{\sqrt{6}%
}(2ssu-sus-uss)$\\\hline
$\Xi^{-}$ & $dss$ & $\frac{1}{\sqrt{2}}(sds-dss)$ & $\frac{1}{\sqrt{6}%
}(2ssd-sds-dds)$\\\hline
$\Omega^{-}$ & $sss$ &  & \\\hline
$\Sigma_{c}$ & $uuc$ & $\frac{1}{\sqrt{2}}(cuu-ucu)$ & $\frac{1}{\sqrt{6}%
}(cuu-ucu-2uuc)$\\\hline
$\Sigma_{b}$ & $uub$ & $\frac{1}{\sqrt{2}}(buu-ubu)$ & $\frac{1}{\sqrt{6}%
}(buu-ubu-2uub)$\\\hline
$\Lambda_{c}$ &  & $\frac{1}{2\sqrt{3}}(ucd+cdu-cud-dcu-2duc+2udc)$ &
$\frac{1}{2}(cud+ucd-cdu-dcu)$\\\hline
$\Lambda_{b}$ &  & $\frac{1}{2\sqrt{3}}(ubd+bdu-bud-dbu-2dub+2udb)$ &
$\frac{1}{2}(bud+ubd-bdu-dbu)$\\\hline
$\Xi_{c}$ & $usc$ & $\frac{1}{\sqrt{2}}(suc-usc)$ & $\frac{1}{\sqrt{6}%
}(2scu-suc-usc)$\\\hline
$\Xi_{b}$ & $usb$ & $\frac{1}{\sqrt{2}}(sub-usb)$ & $\frac{1}{\sqrt{6}%
}(2sbu-sub-usb)$\\\hline
$\Omega_{c}$ & $ssb$ & $\frac{1}{\sqrt{2}}(ssc-scs)$ & $\frac{1}{\sqrt{6}%
}(css-scs-2ssc)$\\\hline
\end{tabular}
$\label{TableKey copy(4)}%
\end{table}%

and the singlet state%
\begin{equation}
\phi^{a}=\frac{1}{\sqrt{6}}(uds+dsu+sud-dus-usd-sdu)
\end{equation}
$\phi$ and $\phi^{\prime}$ combinations are chosen so that for the overall
state (not including the antisymmetric color state) is totally symmetric.
There are eleven possible combinations of these spin and flavor states for
(most of) the known baryons with FSS standing for flavor, spin, and space and
N referring to the SU(3) representation). \ The wave function $\psi_{0}$ is a
total $L=l_{\rho}=l_{\lambda}=0$ wavefunction and $\psi^{\prime}$ and
$\psi^{\prime\prime}$ are total $L=1$ and $l_{\rho}=1$ or $l_{\lambda}=1$
wavefunctions, respectively ($L=1$ states have parity of $-1,$ so $l_{\rho}$
and $l_{\lambda}$ cannot both be 1). \ In addition, $\phi^{\prime}%
,\phi^{\prime\prime},$ $\phi^{s}$ and $\phi^{a}$ are all purely flavor
wavefunctions and $\chi^{\prime},\chi^{\prime\prime},$ $\chi^{s}$ are all
purely spin wavefunctions, having total $S=1/2,1/2$ and $3/2$, respectively.
\ These merely contain all possible combinations of flavor or spin so that the
product of the two gives all possible spin-flavor couplings so that, (not
counting color) the total wave function is symmetric. \ All of these
wavefunctions are orthogonal to the others in the set (that is, $\chi^{\prime
}$ is orthogonal to $\chi^{\prime\prime}$ and $\chi^{s}$ etc.).%

\begin{table}[tbp] \centering
\caption{Total spin-flavor-space wavefunctions}$%
\begin{tabular}
[c]{|l|l|l|l|l|l|}\hline
N & J & L & S & Total State (FSS) & $\Psi$\\\hline
8 & $\frac{1}{2}$ & 0 & $\frac{1}{2}$ & $\frac{1}{\sqrt{2}}(\phi^{\prime}%
\chi^{\prime}+\phi^{\prime\prime}\chi^{\prime\prime})\psi_{0}$ & $\Psi_{1}%
$\\\hline
10 & $\frac{3}{2}$ & 0 & $\frac{3}{2}$ & $\phi^{s}\chi^{s}\psi_{0}$ &
$\Psi_{2}$\\\hline
8 & $\frac{1}{2},\frac{3}{2}$ & 1 & $\frac{1}{2}$ & $\frac{1}{2}[(\phi
^{\prime}\chi^{\prime\prime}+\phi^{\prime\prime}\chi^{\prime})\psi^{\prime
}+(\phi^{\prime}\chi^{\prime}-\phi^{\prime\prime}\chi^{\prime\prime}%
)\psi^{\prime\prime}]$ & $\Psi_{3}(J=\frac{1}{2}),\Psi_{4}(J=\frac{3}{2}%
)$\\\hline
8 & $\frac{1}{2},\frac{3}{2},\frac{5}{2}$ & 1 & $\frac{3}{2}$ & $\frac
{1}{\sqrt{2}}[\phi^{\prime}\chi^{s}\psi^{\prime}+\phi^{\prime\prime}\chi
^{s}\psi^{\prime\prime}]$ & $\Psi_{5}(J=\frac{1}{2}),\Psi_{6}(J=\frac{3}%
{2}),\Psi_{7}(J=\frac{5}{2})$\\\hline
10 & $\frac{1}{2},\frac{3}{2}$ & 1 & $\frac{1}{2}$ & $\frac{1}{\sqrt{2}}%
[\phi^{s}\chi^{\prime}\psi^{\prime}+\phi^{s}\chi^{\prime\prime}\psi
^{\prime\prime}]$ & $\Psi_{8}(J=\frac{1}{2}),\Psi_{9}(J=\frac{3}{2})$\\\hline
1 & $\frac{1}{2},\frac{3}{2}$ & 1 & $\frac{1}{2}$ & $\frac{1}{\sqrt{2}}%
[\phi^{a}\chi^{\prime\prime}\psi^{\prime}-\phi^{a}\chi^{\prime}\psi
^{\prime\prime}]$ & $\Psi_{10}(J=\frac{1}{2}),\Psi_{11}(J=\frac{3}{2}%
)$\\\hline
\end{tabular}
$\label{TableKey copy(5)}%
\end{table}%

\bigskip These wavefunctions then define a grouping of baryons and the
individual baryons themselves are defined by the flavor state from there, as
given below.
\begin{table}[tbp] \centering
\caption{Baryons and their corresponding spin-flavor wavefunctions}%
\begin{align*}
\Psi_{1}  &  \rightarrow p,n,\Lambda,\Sigma^{+},\Sigma^{0},\Sigma^{-},\Xi
^{0},\Xi^{-},N(1440),\Lambda(1600),\Sigma(1660),\Xi(1690),\Sigma_{c}%
^{+}(2455),\Sigma_{b}^{+},\Sigma_{b}^{+},\Lambda_{c}^{+},\Lambda_{c}%
^{+}(2595),\Lambda_{b}^{0}\\
\Psi_{2}  &  \rightarrow\Delta^{++},\Delta^{+},\Delta^{0},\Delta^{-}%
,\Sigma^{+}(1385),\Sigma^{0}(1388),\Sigma^{-}(1390),\Xi^{0}(1530),\Xi
^{-}(1535),\Omega^{-},\Delta(1600),\Sigma(1690)\\
\Psi_{3}  &  \rightarrow N(1535),\Lambda(1670),\Sigma(1750),\Sigma(1880)\\
\Psi_{4}  &  \rightarrow N(1520),\Lambda(1690),\Sigma(1670),\Xi(1820)\\
\Psi_{5}  &  \rightarrow N(1650),\Lambda(1800),\Sigma(1750)\\
\Psi_{6}  &  \rightarrow N(1700),\Sigma(1940)\\
\Psi_{7}  &  \rightarrow N(1675),\Lambda(1830),\Sigma(1775),\Xi(1950)\\
\Psi_{8}  &  \rightarrow\Delta(1620)\\
\Psi_{9}  &  \rightarrow\Delta(1700)\\
\Psi_{10}  &  \rightarrow\Lambda(1405)\\
\Psi_{11}  &  \rightarrow\Lambda(1520)
\end{align*}
\label{TableKey copy(6)}%
\end{table}%
. Before discussing the orbital and radial parts of the wave function we
introduce the coordinate system including relative coordinates for our three
body problem

\ 

\subsection{Coordinate system transforms}

\qquad This section provides a description of how the coordinate system is set
up for the relativistic three-body problem. \ One of the simplest and most
common ways to begin handling a three-body system is to redefine the
coordinate system so that there are only two relative coordinates instead of
three. In the following section we describe the way the coordinate system is
defined for the three-body system and then reduced to two relative
coordinates, plus a "center of mass". We then address how those are further
simplified with additional coordinate transforms in order to analytically
solve as much of the problem as possible before going to numerical methods.

Let us return to Eq. (\ref{duranham}). \ Our goal is to create a coordinate
system in which the kinetic terms can be evaluated analytically and the
variational principle will be used to solve for the energy eigenvalues. \ The
general form of each $\Phi_{ij}$ is%
\begin{align}
\Phi_{ij}  &  =\Phi_{SIij}+\Phi_{Dij}+\mathbf{L\cdot(}\boldsymbol{\sigma}%
_{i}\mathbf{+}\boldsymbol{\sigma}_{j}\mathbf{)}\Phi_{SOij}+\boldsymbol{\sigma
}_{i}\mathbf{\cdot\hat{r}}_{ij}\boldsymbol{\sigma}_{j}\mathbf{\cdot\hat{r}%
}_{ij}\mathbf{L\cdot(}\boldsymbol{\sigma}_{1}\mathbf{+}\boldsymbol{\sigma}%
_{2}\mathbf{)}\Phi_{SOTij}\nonumber\\
&  +\boldsymbol{\sigma}_{i}\mathbf{\cdot}\boldsymbol{\sigma}_{j}%
\Phi_{S\operatorname{Si}j}+(3\boldsymbol{\sigma}_{i}\mathbf{\cdot\hat{r}}%
_{ij}\boldsymbol{\sigma}_{j}\mathbf{\ \cdot\hat{r}}_{ij}\mathbf{-}%
\boldsymbol{\sigma}_{i}\mathbf{\cdot}\boldsymbol{\sigma}_{j})\Phi
_{Tij}+\mathbf{L\cdot(}\boldsymbol{\sigma}_{i}\mathbf{-}\boldsymbol{\sigma
}_{j}\mathbf{)\Phi}_{SODij}, \label{sf}%
\end{align}
\ \ and the various $\Phi$ terms are all functions of $S_{ij,}$ $A_{ij}$
\begin{align}
S_{ij}  &  =\frac{4\pi\Lambda^{2}r_{ij}}{27},\nonumber\\
A_{ij}  &  =-\frac{8\pi}{27r_{ij}\ln(Ke^{2}+\frac{B}{(\Lambda r_{ij})^{2}}%
)}+\frac{e_{1}e_{2}}{4\pi r_{ij}},
\end{align}
and their derivatives (explicit forms given in Appendix A of \cite{whitney}).
\ Note here how they still account for the asymptotic freedom and linear
confinement mentioned earlier. \ The scalar term goes to infinity as $r$ goes
to an infinite value, providing confinement, while the logarithm in the vector
term becomes large at short distance, giving asymptotic freedom (this causes
the vector term to behave like $\sim\alpha/r\ln r)$. \ 

We now define a coordinate system such that in place of the three coordinates
$r_{ij}$ we have two relative coordinates that can be written in terms of the
original $r_{ij}$ distances between each quark pair. \ The notation used is
the same as from \cite{2}, with the actual transformation having individual
particle masses $m_{i}$ replaced by their corresponding c.m. energies
$\varepsilon_{i}.$given in Eq. (\ref{en}). \ A total center of energy system,
$\varepsilon_{1}\mathbf{r}_{1}+\varepsilon_{2}\mathbf{r}_{2}+\varepsilon
_{3}\mathbf{r}_{3}=w\mathbf{R}=0$, has been used to eliminate one of the
coordinates, which is why $\mathbf{r}_{1}$ does not appear in the equations
below for $\mathbf{\rho}$ and $\mathbf{\lambda}$.
\begin{align}
\mathbf{\rho}  &  =\mathbf{r}_{2}-\mathbf{r}_{3},\nonumber\\
\mathbf{\lambda}  &  {\small =}\frac{w\varepsilon_{2}}{(\varepsilon
_{2}+\varepsilon_{3})\varepsilon_{1}}\mathbf{r}_{2}+\frac{w\varepsilon_{3}%
}{(\varepsilon_{2}+\varepsilon_{3})\varepsilon_{1}}\mathbf{r}_{3},\nonumber\\
\mathbf{r}_{1}-\mathbf{r}_{2}  &  =-\frac{\varepsilon_{3}}{\varepsilon
_{2}+\varepsilon_{3}}\mathbf{\rho}-\mathbf{\lambda,}\nonumber\\
\mathbf{r}_{1}-\mathbf{r}_{3}  &  =-\frac{\varepsilon_{2}}{\varepsilon
_{2}+\varepsilon_{3}}\mathbf{\rho}+\mathbf{\lambda,}\nonumber\\
\mathbf{r}_{2}-\mathbf{r}_{3}  &  =\mathbf{\rho,}\nonumber\\
\text{ }\varepsilon_{\rho}  &  =\frac{\varepsilon_{1}(\varepsilon
_{2}+\varepsilon_{3})}{w},\text{ }\varepsilon_{\lambda}=\frac{\varepsilon
_{2}\varepsilon_{3}}{\varepsilon_{2}+\varepsilon_{3}}.
\end{align}
Again, $w$ is the total baryon energy eigenvalue and the epsilons are the
individual c.m. energies of each quark, such that%
\begin{equation}
w=\varepsilon_{1}+\varepsilon_{2}+\varepsilon_{3},
\end{equation}
and the $\varepsilon_{\rho},$ and $\varepsilon_{\lambda}$ can be regarded as
reduced energy terms (similar to reduced mass, but using c.m. energies instead
of masses). \ The corresponding conjugate momenta are given by%
\begin{align}
\mathbf{p}_{\rho}  &  =\frac{\varepsilon_{3}\mathbf{p}_{2}-\varepsilon
_{2}\mathbf{p}_{3}}{\varepsilon_{2}+\varepsilon_{3}},\nonumber\\
\mathbf{p}_{\lambda}  &  =\frac{\varepsilon_{1}}{w}(\mathbf{p}_{2}%
\mathbf{+p}_{3}).
\end{align}
\ In this new system of relative coordinates, the original Hamiltonian of Eq.
(\ref{duranham}) now becomes%
\begin{align}
\mathcal{H}  &  =\frac{1}{F}(\frac{p_{\rho}^{2}}{2\varepsilon_{\rho}%
(E,m_{1},m_{2},m_{3})}+\frac{p_{\lambda}^{2}}{2\varepsilon_{\lambda}%
(E,m_{1},m_{2},m_{3})}\nonumber\\
&  +\frac{\Phi_{12}+\Phi_{13}}{2\varepsilon_{1}(E,m_{1},m_{2},m_{3})}%
+\frac{\Phi_{23}+\Phi_{12}}{2\varepsilon_{2}(E,m_{1},m_{2},m_{3})}+\frac
{\Phi_{31}+\Phi_{23}}{2\varepsilon_{3}(E,m_{1},m_{2},m_{3})}). \label{316}%
\end{align}
\ 

\subsection{Variational Principle and the Gaussian-like Basis Functions}

\qquad Here we will briefly detail how our wavefunctions are used to construct
the basis for use with the variational theorem. \ In order to expand Eq.
(\ref{duranham}) into a general matrix eigenvalue equation, we define
\begin{align}
|\Psi\rangle &  =%
{\displaystyle\sum\limits_{n}}
c_{n}|\Psi_{n}\rangle,\nonumber\\
\langle\Psi|\mathcal{H}|\Psi\rangle &  =%
{\displaystyle\sum\limits_{n,m}}
c_{n}c_{m}^{\ast}\langle\Psi_{m}|\mathcal{H}|\Psi_{n}\rangle=%
{\displaystyle\sum\limits_{n,m}}
c_{n}c_{m}^{\ast}\mathcal{H}_{mn},\nonumber\\
\langle\Psi|\Psi\rangle &  =%
{\displaystyle\sum\limits_{n,m}}
c_{n}c_{m}^{\ast}\langle\Psi_{m}|\Psi_{n}\rangle\equiv%
{\displaystyle\sum\limits_{n,m}}
c_{n}c_{m}^{\ast}B_{mn}. \label{313}%
\end{align}
Since the basis we will choose for the baryons is not orthogonal, $B$ is not
the usual Kronecker delta. \ Using the method of Lagrange multipliers we
arrive at the eigenvalue equation in matrix form (where $\mathbb{H}$ and
$\mathbb{B}$ are matrices, $\mathbf{c}$ is a vector and $E$ our scalar
eigenvalue)%
\begin{equation}
\mathbb{H}\mathbf{c}=E\mathbb{B}\mathbf{c}\text{.} \label{13}%
\end{equation}

As for the radial wavefunctions themselves, we use a Gaussian-like basis
detailed in Appendix A and given by
\begin{align}
&  \frac{u_{l_{\rho}}(\rho)}{\rho}\frac{u_{l_{\lambda}}(\lambda)}{\lambda
};\nonumber\\
\frac{u_{l_{\rho}}(\rho)}{\rho}  &  =\rho^{l_{\rho}}\sum_{n=1}^{2N-1}%
e_{n}\sqrt{a^{3}\sqrt{\frac{f_{n}^{3}}{\pi^{3}}}}\exp(-\frac{f_{n}%
a^{2}\mathbf{\rho}^{2}}{2}),\nonumber\\
\frac{u_{l_{gl}}(\lambda)}{\lambda}  &  =\rho^{l_{\lambda}}\sum_{n=1}%
^{2N-1}e_{n}\sqrt{a^{3}\sqrt{\frac{f_{n}^{3}}{\pi^{3}}}}\exp(-\frac{f_{n}%
a^{2}\mathbf{\lambda}^{2}}{2}),\nonumber\\
f_{n}  &  =\frac{1}{n};~1\leq n\leq N.\nonumber\\
f_{n}  &  =n+1-N;~N+1\leq n\leq2N-1. \label{fn}%
\end{align}
These display an advantage over the usual Gaussian basis in that for a given
choice of the inverse length scale factor $a$ they span both large and small
distances, important when relativistic potentials are included, having broadly
different length scales. Allowing $n_{\rho}$ and $n_{\lambda}$ to stand for
$f_{n}$\ the \ total angular and radial portions for the combined typical wave
function are given by
\begin{equation}
\psi_{n_{\rho}n_{\lambda}l_{\rho}l_{\lambda}LM_{L}}=\sum_{m_{\rho}m_{\lambda}%
}\langle l_{\rho}l_{\lambda}m_{\rho}m_{\lambda}|LM_{L}\rangle\mathcal{N}%
\rho^{l_{\rho}}\lambda^{l_{\lambda}}e^{-n_{\rho}\alpha_{\rho}^{2}\rho
^{2}/2-n_{\lambda}\alpha_{\lambda}^{2}\lambda^{2}/2}Y_{l_{\rho}}^{m_{\rho}%
}Y_{l_{\lambda}}^{m_{\lambda}},
\end{equation}
where $\mathcal{N}$ is a normalization constant. \ \ The general state
$|\Psi\rangle$ and the sum given in Eq. (\ref{313}) (as well as (\ref{314})
below) would \ include the above wave function attached to the appropriate
flavor, color, and spin portions with the index in the summation and
coefficients given in that equation including the summations and $e_{n}$
coefficients given in Eq. (\ref{fn}).

In the two-body problem with tensor coupling as appears in $\Phi_{ij}$ above,
the states $l=j-1$ and $l=j+1$ are mixed, so we need a mixed wavefunction%
\begin{align}
|\Psi\rangle &  =%
{\displaystyle\sum\limits_{n}}
c_{n}^{+}|\Psi_{n+}\rangle+%
{\displaystyle\sum\limits_{n}}
c_{n}^{-}|\Psi_{n-}\rangle,\nonumber\\
-  &  \rightarrow l=j-1,\nonumber\\
+  &  \rightarrow l=j+1.
\end{align}
Using this $\Psi$ in Eq.(\ref{313}) gives
\begin{align}
\langle\Psi|\mathcal{H}|\Psi\rangle &  =%
{\displaystyle\sum\limits_{n,m}}
(c_{m}^{\ast+}c_{n}^{+}\mathcal{H}_{mn}^{++}+c_{m}^{\ast-}c_{n}^{+}%
\mathcal{H}_{mn}^{-+}+c_{m}^{\ast+}c_{n}^{-}\mathcal{H}_{mn}^{+-}+c_{m}%
^{\ast-}c_{n}^{-}\mathcal{H}_{mn}^{--}),\label{314}\\
\langle\Psi|\Psi\rangle &  \equiv%
{\displaystyle\sum\limits_{n,m}}
(c_{m}^{\ast+}c_{n}^{+}B_{mn}^{++}+c_{m}^{\ast-}c_{n}^{-}B_{mn}^{--})\nonumber
\end{align}
in which%
\begin{align}
B_{mn}^{++}  &  =\langle\Psi_{m+}|\Psi_{n+}\rangle,B_{mn}^{--}=\langle
\Psi_{m-}|\Psi_{n-}\rangle,\nonumber\\
\mathcal{H}_{mn}^{++}  &  =\langle\Psi_{m+}|\mathcal{H}|\Psi_{n+}%
\rangle,\mathcal{H}_{mn}^{--}=\langle\Psi_{m-}|\mathcal{H}|\Psi_{n-}%
\rangle,\nonumber\\
\mathcal{H}_{mn}^{+-}  &  =\langle\Psi_{m+}|\mathcal{H}|\Psi_{n-}%
\rangle,\mathcal{H}_{mn}^{-+}=\langle\Psi_{m-}|\mathcal{H}|\Psi_{n+}\rangle.
\end{align}
and similar to before, this leads to the eigenvalue equation with the matrix
structure%
\begin{equation}%
\begin{pmatrix}
\mathbb{H}^{++} & \mathbb{H}^{+-}\\
\mathbb{H}^{-+} & \mathbb{H}^{--}%
\end{pmatrix}%
\begin{pmatrix}
\mathbf{c}^{+}\\
\mathbf{c}^{-}%
\end{pmatrix}
=E%
\begin{pmatrix}
\mathbb{B}^{++} & 0\\
0 & \mathbb{B}^{--}%
\end{pmatrix}%
\begin{pmatrix}
\mathbf{c}^{+}\\
\mathbf{c}^{-}%
\end{pmatrix}
\end{equation}

As outlined above, we use the variational principle%
\begin{equation}
\langle\Psi|\mathcal{H}|\Psi\rangle=E\langle\Psi|\Psi\rangle,
\end{equation}
to find the eigenvalues of our Hamiltonian, with the wavefunction of total $J
$,$L$ and $S$ as
\begin{equation}
|JL(l_{\rho}l_{\lambda})S(S_{1}S_{2}S_{3})Mn_{\rho}n_{\lambda}\rangle
\equiv|\Psi_{n}\rangle.
\end{equation}
where total $L$ is composed of the angular momenta associated with the $\rho$
and $\lambda$ coordinates and total $S$ is composed of the individual spins of
each of the three quarks, with $S_{1},S_{2}$ and $S_{3}$ corresponding to the
spins of quarks 1, 2 and 3, respectively. \ Therefore a typical matrix element
of the Hamiltonian would be%
\begin{equation}
\langle\Psi_{n}|\mathcal{H}|\Psi_{m}\rangle=\langle JL(l_{\rho}l_{\lambda
})S(S_{1}S_{2}S_{3})Mn_{\rho}n_{\lambda}|\mathcal{H}|JL^{\prime}(l_{\rho
}^{\prime}l_{\lambda}^{\prime})S^{\prime}(S_{1}^{\prime}S_{2}^{\prime}%
S_{3}^{\prime})Mn_{\rho}^{\prime}n_{\lambda}^{\prime}\rangle
\end{equation}
\ 

One major advantage of this particular choice of coordinates is that the
kinetic terms can be analytically evaluated. \ The matrix elements for the
kinetic term%

\begin{equation}
\langle\psi_{n_{\rho}n_{\lambda}}|T|\psi_{n_{\rho}^{\prime}n_{\lambda}%
^{\prime}}\rangle=\langle\psi_{n_{\rho}n_{\lambda}}|\frac{1}{F}(\frac{p_{\rho
}^{2}}{2\varepsilon_{\rho}(E,m_{1},m_{2},m_{3})}+\frac{p_{\lambda}^{2}%
}{2\varepsilon_{\lambda}(E,m_{1},m_{2},m_{3})})|\psi_{n_{\rho}^{\prime
}n_{\lambda}^{\prime}}\rangle
\end{equation}
are functions of $l_{\rho},l_{\lambda},n_{\rho},n_{\lambda}$ and the Gaussian
parameters $\alpha_{\rho}$ and $\alpha_{\lambda}$(see Appendix E of
\cite{whitney})

The matrix elements for the potentials, however, must be evaluated
numerically. \ For simplicity, here we just show the matrix elements for the
spin-independent components of the potential (this leaves out the
Clebsch-Gordon coefficients and spherical harmonics since they norm to 1).
\ With the current substitution, the $r_{23}$ term is relatively simple since
$r_{23}=\rho$. \ The matrix elements for the spin-independent $\Phi
_{23}(r_{23})$ quasipotentials ($\Phi_{SI23}(r_{23}))$ and $\Phi_{D23}%
(r_{23})$ ) reduce down to a single radial integral
\begin{align}
&  \langle\psi_{n_{\rho}n_{\lambda}}|\Phi_{23}(\rho)|\psi_{n_{\rho}^{\prime
}n_{\lambda}^{\prime}}\rangle\nonumber\\
&  =\int\Phi_{23}(\rho)\sqrt{\frac{(n_{\rho}\alpha_{\rho}^{2})^{(2l_{\rho
}+3)/2}(n_{\lambda}\alpha_{\lambda}^{2})^{(2l_{\lambda}+3)/2}}{\Gamma
\lbrack(2l_{\rho}+3)/2]\Gamma\lbrack(2l_{\lambda}+3)/2]}}\sqrt{\frac{(n_{\rho
}^{\prime}\alpha_{\rho}^{2})^{(2l_{\rho}^{\prime}+3)/2}(n_{\lambda}^{\prime
}\alpha_{\lambda}^{2})^{(2l_{\lambda}^{\prime}+3)/2}}{\Gamma\lbrack(2l_{\rho
}^{\prime}+3)/2]\Gamma\lbrack(2l_{\lambda}^{\prime}+3)/2]}}\nonumber\\
&  \frac{\Gamma\lbrack(l_{\lambda}+l_{\lambda}^{\prime}+3)/2]}{2[(n_{\lambda
}+n_{\lambda}^{\prime})\alpha_{\lambda}^{2}]^{(l_{\lambda}+l_{\lambda}%
^{\prime}+3)/2}}\rho^{(l_{\rho}+l_{\rho}^{\prime}+2)}e^{-(n_{\rho}+n_{\rho
}^{\prime})\alpha_{\rho}^{2}\rho^{2}/2}d\rho.\nonumber
\end{align}
Thus we are left with a function of one variable that can easily be
numerically integrated regardless of what $\Phi_{23}(\rho)$ happens to be.
\ However, this is not the case for the other two terms. \ 

The matrix elements of the $r_{12}$ and $r_{13}$ spin-independent interactions
are, respectively
\begin{align}
&  \langle\psi_{n_{\rho}n_{\lambda}}|\Phi_{12}(r_{12})|\psi_{n_{\rho}^{\prime
}n_{\lambda}^{\prime}}\rangle\nonumber\\
&  =\int\Phi_{12}(\rho,\lambda)\sqrt{\frac{(n_{\rho}\alpha_{\rho}%
^{2})^{(2l_{\rho}+3)/2}(n_{\lambda}\alpha_{\lambda}^{2})^{(2l_{\lambda}+3)/2}%
}{\Gamma\lbrack(2l_{\rho}+3)/2]\Gamma\lbrack(2l_{\lambda}+3)/2]}}\sqrt
{\frac{(n_{\rho}^{\prime}\alpha_{\rho}^{2})^{(2l_{\rho}^{\prime}%
+3)/2}(n_{\lambda}^{\prime}\alpha_{\lambda}^{2})^{(2l_{\lambda}^{\prime}%
+3)/2}}{\Gamma\lbrack(2l_{\rho}^{\prime}+3)/2]\Gamma\lbrack(2l_{\lambda
}^{\prime}+3)/2]}}\nonumber\\
&  \times\rho^{(l_{\rho}+l_{\rho}^{\prime})}\lambda^{(l_{\lambda}+l_{\lambda
}^{\prime})}e^{-(n_{\rho}+n_{\rho}^{\prime})\alpha_{\rho}^{2}\rho
^{2}/2-(n_{\lambda}+n_{\lambda}^{\prime})\alpha_{\lambda}^{2}\lambda^{2}%
/2}d^{3}\rho d^{3}\lambda,
\end{align}
with a similar expression for $\langle\psi_{n_{\rho}n_{\lambda}}|\Phi
_{13}(r_{13})|\psi_{n_{\rho}^{\prime}n_{\lambda}^{\prime}}\rangle~$and so as
the potentials are now in terms of two variables, it is much more difficult
and time-consuming to numerically evaluate this integral. \ We therefore wish
to make another variable change in the $r_{12}$ and $r_{13}$ systems in order
to rewrite them in terms of a single variable as well.

What now follows is a brief description of the variable change simplification
using the simplest nontrivial case of $l=1$; more explicit and general details
can be found in Appendix E of \cite{whitney}. \ The variable change used is
based on properties of the spherical harmonics and how they relate to
spherical tensors and similarly for the other spherical harmonics. \ We
specialize our discussion to $l=1\ $and use \
\begin{align}
Y_{1}^{0}r  &  =\frac{1}{2}\sqrt{\frac{3}{\pi}}z,\nonumber\\
Y_{1}^{\pm1}r  &  =\mp\frac{1}{2}\sqrt{\frac{3}{2\pi}}(x.\pm iy).
\end{align}
Therefore, since part of our wavefunction is a spherical harmonic (which is a
trigonometric function) and a coordinate, the wavefunction can be rewritten in
spherical tensor form as%
\begin{align}
\Psi_{n}  &  =\frac{4}{3\sqrt{\pi}}n_{\rho}^{5/4}\alpha_{\rho}^{5/2}%
n_{\lambda}^{5/4}\alpha_{\lambda}^{5/2}\rho\lambda e^{-n_{\rho}\alpha_{\rho
}^{2}\rho^{2}/2-n_{\lambda}\alpha_{\lambda}^{2}\lambda^{2}/2}\sum_{m_{\rho
}m_{\lambda}}\langle11m_{\rho}m_{\lambda}|00\rangle Y_{1}^{m_{\rho}}%
Y_{1}^{m_{\lambda}}\nonumber\\
&  =\frac{4}{3\sqrt{\pi}}n_{\rho}^{5/4}\alpha_{\rho}^{5/2}n_{\lambda}%
^{5/4}\alpha_{\lambda}^{5/2}e^{-n_{\rho}\alpha_{\rho}^{2}\rho^{2}%
/2-n_{\lambda}\alpha_{\lambda}^{2}\lambda^{2}/2}\sum_{m_{\rho}m_{\lambda}%
}\langle11m_{\rho}m_{\lambda}|00\rangle\rho_{m_{\rho}}\lambda_{m_{\lambda}},
\end{align}
where%

\begin{align}
\rho_{m_{\rho}}  &  =\rho Y_{1}^{m_{\rho}}(\hat{\rho})\label{firstens}\\
\lambda_{m_{\lambda}}  &  =\lambda Y_{1}^{m_{\lambda}}(\hat{\lambda
}).\nonumber
\end{align}

Additional manipulations are still needed in order to work out the expectation
values explicitly. \ For the $r_{12}$ integration, a new set of variables is
defined as%
\begin{align}
\mathbf{\rho}^{\prime}  &  =\mathbf{r}_{12}=\mathbf{r}_{1}-\mathbf{r}%
_{2},\label{primes}\\
\mathbf{\lambda}^{\prime}  &  =\frac{w\varepsilon_{1}}{\varepsilon
_{3}(\varepsilon_{1}+\varepsilon_{2})}\mathbf{r}_{1}+\frac{w\varepsilon_{2}%
}{\varepsilon_{3}(\varepsilon_{1}+\varepsilon_{2})}\mathbf{r}_{2},\nonumber
\end{align}
and \ then are rewritten in terms of new primed variables and as tensors,
using the same tensor substitution done above%
\begin{align}
\mathbf{\rho}  &  \mathbf{=}\frac{\varepsilon_{1}}{\varepsilon_{1}%
+\varepsilon_{2}}\mathbf{\rho}^{\prime}+\mathbf{\lambda}^{\prime},\nonumber\\
\rho_{m_{\rho}}  &  =\frac{\varepsilon_{1}}{\varepsilon_{1}+\varepsilon_{2}%
}\rho_{m_{\rho}}^{\prime}+\lambda_{m_{\rho}}^{\prime},
\end{align}%
\begin{align}
\mathbf{\lambda}  &  =\frac{w\varepsilon_{2}}{(\varepsilon_{2}+\varepsilon
_{3})(\varepsilon_{1}+\varepsilon_{2})}\mathbf{\rho}^{\prime}-\frac
{\varepsilon_{3}}{\varepsilon_{2}+\varepsilon_{3}}\mathbf{\lambda}^{\prime
},\nonumber\\
\lambda_{m_{\lambda}}  &  =\frac{w\varepsilon_{2}}{(\varepsilon_{2}%
+\varepsilon_{3})(\varepsilon_{1}+\varepsilon_{2})}\rho_{m_{\lambda}}^{\prime
}-\frac{\varepsilon_{3}}{\varepsilon_{2}+\varepsilon_{3}}\lambda_{m_{\lambda}%
}^{\prime}.
\end{align}
Note that this is not a new coordinate system but rather a change of
integration variables. \ This means that, while we are currently working out
the new integral for the $r_{12}$ system, we can use a similar substitution
for the $r_{13}$ system and acquire a nearly identical equation with only a
few constants changed (constants in terms of the integration variable, not
overall constants for the full calculation). \ Before the new substitution of
the primed coordinates, the expectation value of the potential $\Phi
_{12}(\mathbf{\rho}^{\prime}=\mathbf{r}_{12})$ is \
\begin{align}
\langle\psi_{n_{\rho}n_{\lambda}}|\Phi_{12}(\mathbf{r}_{12})|\psi_{n_{\rho
}^{\prime}n_{\lambda}^{\prime}}\rangle &  =\langle\psi_{n_{\rho}n_{\lambda}%
}|\Phi_{12}(\mathbf{\rho}^{\prime})|\psi_{n_{\rho}^{\prime}n_{\lambda}%
^{\prime}}\rangle\nonumber\\
&  =\frac{16}{9\pi}\alpha_{\rho}^{5}\alpha_{\lambda}^{5}(n_{\rho1}%
^{5/4}n_{\lambda1}^{5/4}n_{\rho2}^{5/4}n_{\lambda2}^{5/4})\nonumber\\
&  \times\int\Phi_{12}(\mathbf{\rho}^{\prime})\sum_{m_{\rho1}m_{\lambda1}%
}\langle11m_{\rho1}m_{\lambda1}|00\rangle\rho_{m_{\rho1}}^{\prime\ast}%
\lambda_{m_{\lambda1}}^{\prime\ast}\sum_{m_{\rho2}m_{\lambda2}}\langle
11m_{\rho2}m_{\lambda2}|00\rangle\rho_{m_{\rho2}}^{\prime}\lambda
_{m_{\lambda2}}^{\prime},\nonumber\\
&  \times e^{-(n_{\rho1}+n_{\rho2})\alpha_{\rho}^{2}\rho^{\prime2}%
-(n_{\lambda1}+n_{\lambda2})\alpha_{\lambda}^{2}\lambda^{2}}d^{3}\rho^{\prime
}d^{3}\lambda^{\prime}.
\end{align}
We use the derived relationships between the primed and un-primed coordinates
$\rho$ and $\lambda$ Eq. (\ref{primes}) and one final coordinate change to
eliminatel $\mathbf{\lambda}^{\prime}\mathbf{\cdot\rho}^{\prime}$ cross terms
in the Gaussian. The explicit details can be found in Appendix E of
\cite{whitney}. \ The end result is \ that $\langle\psi_{n_{\rho}n_{\lambda}%
}|\Phi_{12}(\mathbf{r}_{12})|\psi_{n_{\rho}^{\prime}n_{\lambda}^{\prime}%
}\rangle$ involves a single radial integral which can be numerically evaluated
easily. \ Similarly the matrix element $\langle\psi_{n_{\rho}n_{\lambda}}%
|\Phi_{13}(\mathbf{r}_{13})|\psi_{n_{\rho}^{\prime}n_{\lambda}^{\prime}%
}\rangle$ can be evaluated. Again, details are listed in Appendix E of
\cite{whitney}. \ The potential is now in terms of just one variable, so
regardless of what potential is used, the numerical calculations will be
fairly straightforward. \ Thus, the coordinate system has been defined and
transformed in such a way as to make a good deal of the problem analytic while
keeping what is not analytic still easy to evaluate numerically. With the
matrix elements defined for a general potential and for analytic kinetic
terms, we now need to explicitly define our potential model. \ 

\section{Three body spin-dependent potentials}

\qquad Conceptually speaking, the approach one would take to go from a
two-body system with the formalism we have described to a three body one is
straightforward. \ The problem is now treated as three\ two-body problems,
with the overall form of the potentials given in Eq. (\ref{sf}). \ The
three-body potential is of similar form and essentially just triples the
number of terms, with pairwise interactions for all three quarks. \ For the
relatively simple vector $(A)$, scalar $(S)$ and Darwin $(\Phi_{D})$ terms
this is almost trivial, as there are no direct spin-dependent operators;
however, the spin-spin, tensor, and spin-orbit terms require extensive
reworking, which are outlined in the following sections with details in
Appendix F of \cite{whitney}.

\subsection{State Couplings and Operator Methods}

\qquad We will now describe how we set up our three-body states when using the
spin-dependent potential operators. \ In order to simplify our numerical
calculations, it is helpful to note that the potential terms are products of a
term involving the coupled angular momentum operators and coordinate dependent
terms that have trivial operator dependence, save for the Clebsch-Gordon
coefficients and spherical harmonics not norming to 1 in the cases where we
have orbital dependence in the operator. \ Even in this case though, the
results of the preceding section still can be applied directly with the added
component. \ This allows us to use the operator angular momentum on a
specified state and just get a number back that depends on the angular
components of the state itself and not any radial components, so that the
numerical integral itself does not involve any angular momentum
operators.\ Thus, our potential terms separated into operator and non-operator
pieces are given in table 4
\begin{table}[tbp] \centering
\caption{Potential terms, operators and non-operator components}$%
\begin{tabular}
[c]{|l|l|l|}\hline
Potential Term & Angular Momenta Operator components & Non-operator
component\\\hline
Spin-Spin & $\boldsymbol{\sigma}_{i}\mathbf{\cdot}\boldsymbol{\sigma}_{j}$ &
$\Phi_{SS}(\mathbf{r}_{ij}\mathbf{)}$\\\hline
Spin-Orbit & $\mathbf{L}_{ij}\mathbf{\cdot(}\boldsymbol{\sigma}_{i}%
\mathbf{+}\boldsymbol{\sigma}_{j}\mathbf{)}$ & $\Phi_{SO}(\mathbf{r}%
_{ij}\mathbf{)} $\\\hline
Spin-Orbit Difference & $\mathbf{L}_{ij}\mathbf{\cdot(}\boldsymbol{\sigma}%
_{i}\mathbf{-}\boldsymbol{\sigma}_{j}\mathbf{)}$ & $\Phi_{SOD}(\mathbf{r}%
_{ij}\mathbf{)}$\\\hline
Tensor & $3\boldsymbol{\sigma}_{i}\mathbf{\cdot\hat{r}}_{ij}\boldsymbol{\sigma
}_{j}\mathbf{\ \cdot\hat{r}}_{ij}\mathbf{-}\boldsymbol{\sigma}_{i}%
\mathbf{\cdot}\boldsymbol{\sigma}_{j}$ & $\Phi_{T}(\mathbf{r}_{ij}\mathbf{)}%
$\\\hline
Spin-Orbit Cross & $i\mathbf{L}_{ij}\mathbf{\cdot}\boldsymbol{\sigma}%
_{i}\mathbf{\times}\boldsymbol{\sigma}_{j}$ & $\Phi_{SOX}(\mathbf{r}%
_{ij}\mathbf{)}$\\\hline
Spin-Orbit Tensor & $\boldsymbol{\sigma}_{i}\mathbf{\cdot\hat{r}}%
_{ij}\boldsymbol{\sigma}_{j}\mathbf{\cdot\hat{r}}_{ij}\mathbf{L}%
_{ij}\mathbf{\cdot(}\boldsymbol{\sigma}_{i}\mathbf{+}\boldsymbol{\sigma}%
_{j}\mathbf{)}$ & $\Phi_{SOT}(\mathbf{r}_{ij}\mathbf{)}$\\\hline
\end{tabular}
$\label{TableKey copy(8)}%
\end{table}%
and the explicit forms of the $\Phi$ terms are given in Appendix A of
\cite{whitney} and the results of all these operators on the possible baryon
configurations are given in Appendix F of \cite{whitney}. \ The above
operators do not affect the radial part of the wavefunction and so the problem
is broken into a radial integral part (as done in the previous section) and an
operator component for each interaction term.

For the baryons we have considered, there are a total of eleven different
wavefunctions, which represent all possible spin-flavor couplings for the
various particles. \ The form of these is given in table 2 where it has been
split into three components: spin, flavor and space\ (represented by
$\chi,\phi,\psi,$ respectively), explicitly defined in Appendix F of
\cite{whitney}. \ As there are six interactions to consider and three
couplings per interaction (we are using two-body operators, so there is a 1-2,
1-3, and 2-3 term for each operator), there are a total of 198 possible
interactions to consider. \ Fortunately, many of these are similar or trivial
and so the number that must actually be worked out explicitly drops
considerably, but there still are a quite a large number that are non-trivial.
\ The eleven wavefunctions are given\ in table 2. \ Explicit forms of these
terms are given in Appendix F of \cite{whitney}. \ The quark flavor
combination of the $\phi$ terms is different for each baryon, but since the
operators we use do not affect the flavor, it does not matter what they are
for the purposes of calculating the effects of each operator.\ 

There are two methods we use to determine the affect of these operators. One
is a simple ladder operator approach and the other involves use of the Wigner
6j and 9j recoupling coefficients, the details of which are given in Appendix
F of \cite{whitney}. \ Both methods are always valid, but not necessarily
always useful due to how the operator form affects each individual
wavefunction for the ladder operators. \ It is worth noting that having two
methods be viable also allows for a good check. \ The ladder operator method
works out simply for all operators (except the Spin-Orbit Cross term due to
the matrix elements being independent of total $M)$. \ For the states
$\Psi_{1},\Psi_{2},\Psi_{4},\Psi_{7},\Psi_{9}$ and $\Psi_{11},$ we can set
$M=J$ and force $M_{s}=S$. This means that any operator that changes total
$M_{s}$ will be orthogonal to the original wavefunction and thus we can
eliminate any term that does change total $M_{s}.$ \ All of the methods for
determining these states are relegated to Appendix F of \cite{whitney}, this
includes the ladder operators, 6j and 9j details. \ Due to having matrix
elements for the two-body problem already defined by \cite{becker}, the
difficult part of this problem is recoupling the state into one which can use
these matrix elements.\ To summarize this section, we have written the
three-body potentials in terms of two relative coordinates and shown how they
can be transformed for each interacting pair. \ This allows a description of
the methods used to adapt the (\cite{jim,wongyoon,alstine2009}) two-body
potential operators of Crater et al. derived for the meson spectrum to the
three-body problem.

\section{\bigskip Numerical Results and Comments}

\qquad The expectation value of the Hamiltonian in Eq. (\ref{duranham}) cannot
be evaluated analytically, so it falls to numerical studies to acquire an
explicit number. \ We use a Monte Carlo approach combined with a simple
gradient method to obtain a best-fit $\chi^{2}$ for the spectrum as a whole,
as compared to current experimental data. \ It is important to note that a
normal $\chi^{2}$ routine would include in each individual baryon's
contribution to the by the inverse square of the experimental error. \ But
this would give particles such as the proton a much higher weight than desired
in the overall fit. \ Therefore, we instead divided each by the greater of
their respective experimental errors or 1 MeV, thus preventing very well-known
particles from dominating our imperfect fit. \ The following sections describe
the numerical methods used and give the results after using said
methods.\bigskip

\ 

\subsection{Methods and Parameter Values}

The numerical best fits were done using a Monte Carlo approach followed by a
gradient method to obtain a least square fit for the spectrum as a whole. \ We
originally attempted to use a more simplified gradient approach but it quickly
became apparent that the functions are far too sensitive to changes and thus
would get "stuck" in a local minimum much too easily without some other
approach. \ So, we adopted a Monte Carlo routine that would trigger whenever
the gradient approach found a new best fit in order to ensure we were reaching
the best results for our theory. \ The integrations were done numerically
using Gaussian Quadrature and the parameters $\alpha_{\rho}$ and
$\alpha_{\lambda}$ were minimized by the Nelder-Mead simplex method, though it
is worth noting that the $\alpha$ parameters do not generally vary much from
the analytic result if one were to use a harmonic oscillator model. Also note
that as the size of our matrix increases, the actual value of these parameters
do not affect the fit as much, becoming irrelevant at an infinitely large
matrix. \ As one might expect, benefits from increasing the size of the matrix
are subject to diminishing returns and thus our results are given for a point
of reasonable convergence (in other words, once increasing the size of the
matrix no longer significantly affected results).

Our model has a total of 8 parameters,
\begin{table}[tbp] \centering
\caption{Parameter values}$%
\begin{tabular}
[c]{|c|c|}\hline
& This work\\\hline
u & $157.2$ (MeV)\\\hline
d & $158.3$ (MeV)\\\hline
s & $337.5$ (MeV)\\\hline
$\Lambda$ & $285.8$ (MeV)\\\hline
c & $2050.3$ (MeV)\\\hline
b & $5302.5$ (MeV)\\\hline
K & $18.1$\\\hline
B & $100.6$\\\hline
\end{tabular}%
\begin{tabular}
[c]{|c|c|}\hline
& Reference 2\\\hline
u & $55.7~$(MeV)\\\hline
d & $55.3$ (MeV)\\\hline
s & $249.9$ (MeV)\\\hline
$\Lambda$ & $421.8~$(MeV)\\\hline
c & $1.476~$ (GeV)\\\hline
b & $4.844$(GeV)\\\hline
K & $4.198$\\\hline
B & $0.05081$\\\hline
\end{tabular}
\ $\label{TableKey}%
\end{table}
with u,d,s,c and b in the table corresponding to the masses of the up, down,
strange, charm and bottom quarks, respectively, $\ $and $\Lambda$, $K$\ and
$B$ are coupling constants in our model. These are the same 8 parameters as in
\cite{wongyoon}.\ It is worth noting that our model has significantly fewer
parameters than most models, with only 8 total and 5 of those being universal
to any model (the quark masses themselves)\footnote{\cite{2}, for example, has
14 parameters listed.}. \ The model of course, is still expected to be
accurate regardless of the number of parameters, but it is worth noting in
this work. \ In addition, there are only two parametric functions that define
our potential model, the vector and scalar potentials $A(r)$ and $S(r)$ given
by Eq.(\ref{sa}). \ 

\subsection{Results and Comparison to\ Experiment}

The complete results of our model are given in tables 5-8. \ As the purpose of
this work is to test if the model used in the two-body case works well for the
three-body, we are only using those baryons which have a three or four star
rating by the Particle Data Group, meaning that they are fairly well-known. \
\begin{table}[tbp] \centering
\caption{Low lying baryon states}$%
\begin{tabular}
[c]{|c|c|c|c|c|c|c|}\hline
Baryon & $J$ & $L$ & $S$ & Theoretical Mass (MeV) & Experimental Mass(MeV) &
Exp-Theory(MeV)\\\hline
$p$ & 1/2 & 0 & 1/2 & 947 & 938 & -9\\\hline
$n$ & 1/2 & 0 & 1/2 & 948 & 939 & -9\\\hline
$\Sigma^{+}$ & 1/2 & 0 & 1/2 & 1250 & 1189 & -61\\\hline
$\Sigma^{0}$ & 1/2 & 0 & 1/2 & 1261 & 1192 & -68\\\hline
$\Sigma^{-}$ & 1/2 & 0 & 1/2 & 1271 & 1197 & -73\\\hline
$\Xi^{0}$ & 1/2 & 0 & 1/2 & 1373 & 1314 & -58\\\hline
$\Xi^{-}$ & 1/2 & 0 & 1/2 & 1378 & 1321 & -57\\\hline
$\Lambda^{0}$ & 1/2 & 0 & 1/2 & 1082 & 1125 & 43\\\hline
$\Delta^{++}$ & 3/2 & 0 & 3/2 & 1249 & 1232 & -17\\\hline
$\Delta^{+}$ & 3/2 & 0 & 3/2 & 1250 & 1232 & -18\\\hline
$\Delta^{0}$ & 3/2 & 0 & 3/2 & 1251 & 1232 & -19\\\hline
$\Delta^{-}$ & 3/2 & 0 & 3/2 & 1252 & 1232 & -20\\\hline
$\Sigma^{+}(1390)$ & 3/2 & 0 & 3/2 & 1384 & 1383 & -1\\\hline
$\Sigma^{0}(1390)$ & 3/2 & 0 & 3/2 & 1385 & 1384 & -1\\\hline
$\Sigma^{-}(1390)$ & 3/2 & 0 & 3/2 & 1387 & 1387 & 0\\\hline
$\Xi^{0}(1530)$ & 3/2 & 0 & 3/2 & 1501 & 1531 & 30\\\hline
$\Xi^{-}(1530)$ & 3/2 & 0 & 3/2 & 1507 & 1535 & 28\\\hline
$\Omega^{-}$ & 3/2 & 0 & 3/2 & 1609 & 1672 & 63\\\hline
\end{tabular}
$\label{TableKey copy(1)}%
\end{table}%

The lowest lying baryons are generally slightly high energy-wise for the first
8 and this is most likely to allow the following 10 to be fit relatively
accurately. \ This is not a surprising result of our model due to the fact
that since we are using no purely 3-body potentials, the only difference
between these sets of baryons is the spin-spin interaction. \ The fitting
routine used the average value for the experimental masses given in table 7
since these have a wide range , and this value is also used in calculating the
difference between our model and experimental data. \ \ In this table the
first 6 listed baryons are radial excitations of ones in the previous table.%

\begin{table}[tbp] \centering
\caption{Orbital and radially excited  baryons states}$%
\begin{tabular}
[c]{|c|c|c|c|c|c|c|}\hline
Baryon & $J$ & $L$ & $S$ & Theoretical Mass (MeV) & Experimental Mass(MeV) &
Exp-Theory\\\hline
N(1440) & 1/2 & 0 & 1/2 & 1557 & 1420-1470 & -117\\\hline
$\Lambda(1600)$ & 1/2 & 0 & 1/2 & 1677 & 1560-1700 & -77\\\hline
$\Sigma(1660)$ & 1/2 & 0 & 1/2 & 1672 & 1630-1690 & 12\\\hline
$\Xi(1690)$ & 1/2 & 0 & 1/2 & 1784 & 1680-1700 & -94\\\hline
$\Delta(1600)$ & 3/2 & 0 & 3/2 & 1521 & 1550-1700 & 78\\\hline
$\Sigma(1670)$ & 3/2 & 1 & 1/2 & 1679 & 1665-1685 & -4\\\hline
N(1535) & 1/2 & 1 & 1/2 & 1549 & 1525-1545 & -14\\\hline
$\Lambda(1670)$ & 1/2 & 1 & 1/2 & 1671 & 1660-1680 & -1\\\hline
$\Sigma(1750)$ & 1/2 & 1 & 3/2 & 1644 & 1730-1800 & 121\\\hline
$\Sigma(1775)$ & 5/2 & 1 & 3/2 & 1661 & 1770-1780 & 114\\\hline
N(1520) & 3/2 & 1 & 1/2 & 1551 & 1515-1525 & -31\\\hline
$\Lambda(1690)$ & 3/2 & 1 & 1/2 & 1670 & 1685-1695 & 20\\\hline
$\Xi(1820)$ & 3/2 & 1 & 1/2 & 1777 & 1818-1828 & 43\\\hline
N(1650) & 1/2 & 1 & 3/2 & 1566 & 1645-1670 & 84\\\hline
$\Lambda(1800)$ & 1/2 & 1 & 3/2 & 1658 & 1720-1850 & 142\\\hline
$\Sigma(1880)$ & 1/2 & 0 & 1/2 & 1709 & 1800-1960 & 171\\\hline
N(1700) & 3/2 & 1 & 3/2 & 1568 & 1650-1750 & 132\\\hline
N(1675) & 5/2 & 1 & 3/2 & 1615 & 1670-1680 & 59\\\hline
$\Lambda(1830)$ & 5/2 & 1 & 3/2 & 1641 & 1810-1830 & 189\\\hline
$\Xi(1950)$ & 5/2 & 1 & 3/2 & 1757 & 1935-1965 & 192\\\hline
$\Delta(1620)$ & 1/2 & 1 & 1/2 & 1542 & 1600-1660 & 78\\\hline
$\Delta(1700)$ & 3/2 & 1 & 1/2 & 1546 & 1670-1750 & 154\\\hline
$\Lambda(1405)$ & 1/2 & 1 & 1/2 & 1410 & 1402-1410 & -4\\\hline
$\Lambda(1520)$ & 3/2 & 1 & 1/2 & 1680 & 1518-1521 & -160\\\hline
\end{tabular}
$\label{TableKey copy(2)}%
\end{table}%

The higher order baryons fall within an acceptable range on the whole, though
there are a few outliers. \ Of important note is that our model does fit very
well the often troublesome $\Lambda(1405)$ particle. \ The other $\Lambda$
particles however are, as before, missing some sort of interaction that will
aid in differentiating among them (the $\Lambda(1520)$ $\Lambda(1670),\Lambda
(1690)$ and $\Lambda(1800)$ all fit to around the same value). \ 

In addition, we fit a the well-known charmed and bottom baryons, given in
table 8. \ The orbital and spin angular momenta are the same as the
non-charmed/bottom baryons that correspond to each charmed or bottom baryon
here.
\begin{table}[tbp] \centering
\caption{Charmed and bottom baryons}$%
\begin{tabular}
[c]{|c|c|c|c|c|c|c|}\hline
Baryon & $J$ & $L$ & $S$ & Theoretical Mass (MeV) & Experimental Mass(MeV) &
Exp-Theory(MeV)\\\hline
$\Sigma_{c}^{++}(2455)$ & 1/2 & 0 & 1/2 & 2385 & 2454 & 68\\\hline
$\Sigma_{c}^{++}(2520)$ & 3/2 & 0 & 3/2 & 2551 & 2520 & -31\\\hline
$\Lambda_{c}^{+}(2286)$ & 1/2 & 0 & 1/2 & 2382 & 2286 & -96\\\hline
$\Lambda_{c}^{+}(2595)$ & 1/2 & 1 & 1/2 & 2415 & 2595 & 180\\\hline
$\Xi_{c}^{+}(2467)$ & 1/2 & 0 & 1/2 & 2561 & 2467 & -94\\\hline
$\Xi_{c}^{0}(2470)$ & 1/2 & 0 & 1/2 & 2562 & 2470 & -92\\\hline
$\Xi_{c}^{+}(2645)$ & 3/2 & 0 & 3/2 & 2598 & 2645 & 46\\\hline
$\Xi_{c}^{+}(2790)$ & 1/2 & 1 & 3/2 & 2661 & 2790 & 129\\\hline
$\Xi_{c}^{+}(2815)$ & 3/2 & 1 & 3/2 & 2707 & 2815 & 108\\\hline
$\Omega_{c}^{0}(2695)$ & 1/2 & 0 & 1/2 & 2732 & 2695 & -37\\\hline
$\Omega_{c}^{0}(2770)$ & 3/2 & 0 & 3/2 & 2745 & 2770 & 25\\\hline
$\Sigma_{b}^{+}(5829)$ & 3/2 & 0 & 3/2 & 5800 & 5829 & 29\\\hline
$\Sigma_{b}^{-}(5836)$ & 3/2 & 0 & 3/2 & 5851 & 5836 & -15\\\hline
$\Xi_{b}^{0}(5790)$ & 1/2 & 0 & 1/2 & 5854 & 5790 & -64\\\hline
$\Omega_{b}^{-}(6071)$ & 1/2 & 0 & 1/2 & 6032 & 6071 & 39\\\hline
\end{tabular}
$\label{TableKey copy(3)}%
\end{table}
These agree relatively well with experimental data.

\qquad

\subsection{Conclusion and Future Work}

\qquad The model has shown that with the use of three-body equation of
Sazdjian, it is possible to use the purely two-body approach based on Dirac's
constraint dynamics for spin-one-half particle bound states for a good fit of
the baryon spectrum. \ As for future work, one may try to see the effects of
higher order eigenvalue equations for the three body system, as referred to in
Appendix B of \cite{whitney} and as discussed in more detail in \cite{5}. It
may be also be possible as in \cite{2} to introduce three-body forces in
addition to the two-body ones and to use a fully three-body approach for a
coordinate system and $JLS$ couplings. \ Total $JLS $ couplings for a
three-body system are usually done in a mathematically rigorous fashion by
coupling two particles together and then coupling their Clebsch-Gordon coupled
two-body system to a third particle for a complete three-body system. \ A
fully three-body approach(\cite{3angmom}) to angular momentum couplings may at
the very least yield a more elegant formalism and perhaps better overall
results. \ A system derived purely for a three-body problem and including
three-body $JLS$\ couplings may include additional interactions not seen in a
two-body model. \ We believe this may solve the issue of the same family of
particles (i.e. $\Sigma,\Lambda,N)$ lacking in enough differentiation as one
goes from one $J$ to another, since the angular momentum dependent
interactions are the only things that accounts for the difference in mass
among different sets of baryons with the same quark configuration. \ In
contrast to many other models(\cite{capstickreview}) which tend to fit the
lower mass baryons very well and the higher order much more poorly, our work
tends to maintain the same quality of fit regardless of baryon mass.\ This
lends credence to the theory as a whole being fundamentally sound, but merely
incomplete. \ This missing piece is likely a higher order implementation of
Sazdjian's three body scheme and/or the fully three-body interactions that
were not considered in this work; three-body interactions referring to those
in which an interaction between two of the particles can influence the third
(as done in \cite{2}), rather than being entirely based on two-body
interactions. \ As can be seen from the fit data, there are many lower than
experiment and many higher as well, though this is spread out among all the
baryons with the low-lying baryons being larger than experiment while the
higher order and charmed/bottom baryons are lower. \ This prevents one from
improving the fits to the low-lying baryons by simply lowering the $u,d,s$
masses since simultaneously this would worsen in the fits on the already low
charmed/bottom baryons (both sets of baryons have similar experimental
errors). \ 

On the whole though, the fit is nearly as accurate as others, most notably the
work of Capstick and Isgur (\cite{2}), which is generally regarded as one of
the more valuable references for theoretical baryon spectroscopy. \ The only
marked difference of the results of our model versus other models is that the
quality of the fit remains relatively constant regardless of which baryons we
are considering (ground state, higher order, heavy, etc.). \ However, as was
discussed, this may actually reinforce that the fundamental approach is sound
and it can be upgraded to a more accurate model by considering additional
interactions and a more refined treatment of Sazdjian's approach to the 3-body
problem of bound systems.

\bigskip

\ 

\section{Appendix}

\appendix%
\makeatletter\@addtoreset{equation}{section}
\makeatother\renewcommand{\theequation}{\Alph{section}\arabic{equation}}%

\section{ \ \ \bigskip Gaussian wavefunctions and Infinite Interval
Discretization}

\qquad This section describes how our wavefunctions comes about for our basis.
\ The potentials in Eq. (\ref{duranham}) have both short distance and long
distance effects, so we need a basis wavefunction that can accurately account
for that. \ We define a wavefunction in terms of some parameter $\alpha$ that
determines the effect of the Gaussian wavefunction for short and long distance
interactions. \ We then split the wavefunction into those two parts\ (short
and long) and discretize it to a certain $N$ value, from which we get our
basis. \ The wavefunction is originally defined in an infinite vector space,
so we must truncate it in order to work with it. \ 

\bigskip Boris Kupershmidt, a mathematician,\cite{Boris} has suggested a
Laplace transform/Gaussian basis
\begin{equation}
\psi(\mathbf{x})=\int_{0}^{\infty}d\alpha q(\alpha)\sqrt{\frac{1}{a^{3}}%
\sqrt{\frac{\alpha^{3}}{\pi^{3}}}}\exp(-\frac{\alpha\mathbf{x}^{2}}{2a^{2}})
\end{equation}
where $\psi$ is essentially the Fourier transform of some function
$q(\alpha).$ \ In order to work with this function, we split the integral into
two pieces, one with boundaries from zero to one and and the other with
boundaries from one to infinity, so that%

\begin{align}
\psi(\mathbf{x})  &  =\int_{0}^{1}d\alpha q(\alpha)\sqrt{\frac{1}{a^{3}}%
\sqrt{\frac{\alpha^{3}}{\pi^{3}}}}\exp(-\frac{\alpha\mathbf{x}^{2}}{2a^{2}%
})\nonumber\\
&  +\int_{1}^{\infty}d\alpha q(\alpha)\sqrt{\frac{1}{a^{3}}\sqrt{\frac
{\alpha^{3}}{\pi^{3}}}}\exp(-\frac{\alpha\mathbf{x}^{2}}{2a^{2}}).
\end{align}
By replacing $\alpha$ with $1/\beta$ in the first half of the equation (so
that the integral from 0 to 1 now becomes 1 to infinity) we get%
\begin{align}
\psi(\mathbf{x})  &  =\int_{1}^{\infty}d\beta q(1/\beta)\sqrt{\frac{1}{a^{3}%
}\sqrt{\frac{1}{\beta^{3}\pi^{3}}}}\exp(-\frac{\mathbf{x}^{2}}{2\beta a^{2}%
})\nonumber\\
&  +\int_{1}^{\infty}d\alpha q(\alpha)\sqrt{\frac{1}{a^{3}}\sqrt{\frac
{\alpha^{3}}{\pi^{3}}}}\exp(-\frac{\alpha\mathbf{x}^{2}}{2a^{2}}),
\end{align}
and from there, replacing integrals with sums over arbitrarily large $N$, this
discretizes to%
\begin{align}
\psi(\mathbf{x})  &  =\sum_{n=1}^{N}c_{n}\sqrt{\frac{1}{a^{3}}\sqrt{\frac
{1}{n^{3}\pi^{3}}}}\exp(-\frac{\mathbf{x}^{2}}{2na^{2}})\nonumber\\
&  +\sum_{n=1}^{N}d_{n}\sqrt{\frac{1}{a^{3}}\sqrt{\frac{n^{3}}{\pi^{3}}}}%
\exp(-\frac{n\mathbf{x}^{2}}{2a^{2}})\nonumber\\
&  =e_{1}\sqrt{\frac{1}{a^{3}}\sqrt{\frac{1}{\pi^{3}}}}\exp(-\frac
{\mathbf{x}^{2}}{2a^{2}})\nonumber\\
&  +\sum_{n=2}^{N}c_{n}\sqrt{\frac{1}{a^{3}}\sqrt{\frac{1}{n^{3}\pi^{3}}}}%
\exp(-\frac{\mathbf{x}^{2}}{2na^{2}})+\sum_{n=2}^{N}d_{n}\sqrt{\frac{1}{a^{3}%
}\sqrt{\frac{n^{3}}{\pi^{3}}}}\exp(-\frac{n\mathbf{x}^{2}}{2a^{2}}).
\end{align}
So, for $N=1$ we have
\begin{equation}
\psi(\mathbf{x})=e_{1}\sqrt{\frac{1}{a^{3}}\sqrt{\frac{1}{\pi^{3}}}}%
\exp(-\frac{\mathbf{x}^{2}}{2a^{2}}),
\end{equation}
For $N=2$ we have%
\begin{align}
\psi(\mathbf{x})  &  =e_{1}\sqrt{\frac{1}{a^{3}}\sqrt{\frac{1}{\pi^{3}}}}%
\exp(-\frac{\mathbf{x}^{2}}{2a^{2}})\nonumber\\
&  +c_{2}\sqrt{\frac{1}{a^{3}}\sqrt{\frac{1}{8\pi^{3}}}}\exp(-\frac
{\mathbf{x}^{2}}{4a^{2}})+d_{2}\sqrt{\frac{1}{a^{3}}\sqrt{\frac{8}{\pi^{3}}}%
}\exp(-\frac{\mathbf{x}^{2}}{a^{2}})
\end{align}
Note that the original wavefunction from $N=1$ remains as the first term.
\ This is true for all $N.$

For $N\geq3$%
\begin{align}
\psi(\mathbf{x})  &  =e_{1}\sqrt{\frac{1}{a^{3}}\sqrt{\frac{1}{\pi^{3}}}}%
\exp(-\frac{\mathbf{x}^{2}}{2a^{2}})\nonumber\\
&  +\sum_{n=2}^{N}c_{n}\sqrt{\frac{1}{a^{3}}\sqrt{\frac{1}{n^{3}\pi^{3}}}}%
\exp(-\frac{\mathbf{x}^{2}}{2na^{2}})+\sum_{n=2}^{N}d_{n}\sqrt{\frac{1}{a^{3}%
}\sqrt{\frac{n^{3}}{\pi^{3}}}}\exp(-\frac{n\mathbf{x}^{2}}{2a^{2}})
\end{align}
or more symmetrically%
\begin{align}
\psi(\mathbf{x})  &  =\sum_{n=1}^{2N-1}e_{n}\sqrt{\frac{1}{a^{3}}\sqrt
{\frac{f_{n}^{3}}{\pi^{3}}}}\exp(-\frac{f_{n}\mathbf{x}^{2}}{2a^{2}%
}),\nonumber\\
f_{n}  &  =\frac{1}{n};~1\leq n\leq N;~\nonumber\\
f_{n}  &  =n+1-N;~N+1\leq n\leq2N-1
\end{align}
\ As we can see from the $N=2$ case, the order of the matrix increases as
$2N-1$. \ Each matrix element of the Hamiltonian matrix is constructed from
the expectation value of the Hamiltonian with two of these wavefunctions.
\ For example, for the $N=2$ case, our general wavefunction $|\psi_{n}%
(f_{n})\rangle$ is%
\begin{align}
n  &  =1\rightarrow|\psi_{1}(1)\rangle,\nonumber\\
n  &  =2\rightarrow|\psi_{2}(\frac{1}{2})\rangle,\nonumber\\
n  &  =3\rightarrow|\psi_{3}(2)\rangle,
\end{align}
and thus we have the 3x3 matrix%
\begin{equation}%
\begin{pmatrix}
\langle\psi_{1}|\mathcal{H}|\psi_{1}\rangle & \langle\psi_{1}|\mathcal{H}%
|\psi_{2}\rangle & \langle\psi_{1}|\mathcal{H}|\psi_{3}\rangle\\
\langle\psi_{2}|\mathcal{H}|\psi_{1}\rangle & \langle\psi_{2}|\mathcal{H}%
|\psi_{2}\rangle & \langle\psi_{2}|\mathcal{H}|\psi_{3}\rangle\\
\langle\psi_{3}|\mathcal{H}|\psi_{1}\rangle & \langle\psi_{3}|\mathcal{H}%
|\psi_{2}\rangle & \langle\psi_{3}|\mathcal{H}|\psi_{3}\rangle
\end{pmatrix}
.
\end{equation}
As can be inferred from the values of $f_{n}$ for $n>1,$ this basis allows the
wavefunction to account for both the long-rage and short-range interactions of
the Hamiltonian. \ Smaller $f_{n}$ values-such as for $n=2$ in the above
example-allow for long-range interactions while larger $f_{n}$ values (like
the $n=3$ wavefunction) account for the short-range interactions.

In a similar manner, we can now also write our $B$ matrix from Eq.(\ref{13})
as%
\begin{align}
\psi(\mathbf{x})  &  =\sum_{n=1}^{2N-1}e_{n}\psi_{n}(\mathbf{x);}\nonumber\\
B_{nm}  &  =\int d^{3}x\psi_{n}^{\ast}(\mathbf{x)}\psi_{m}(\mathbf{x)}%
\nonumber\\
&  =\sqrt{\frac{1}{a^{6}}\sqrt{\frac{f_{n}^{3}f_{m}^{3}}{\pi^{6}}}}\int
d^{3}x\exp(-\frac{(f_{n}+f_{m})\mathbf{x}^{2}}{2a^{2}})\nonumber\\
&  =\sqrt{f_{n}^{3/2}f_{m}^{3/2}}\frac{\sqrt{8}}{(f_{n}+f_{m})^{3/2}}%
=\sqrt{\frac{8f_{n}^{3/2}f_{m}^{3/2}}{(f_{n}+f_{m})^{3}}}%
\end{align}
Thus we get an analytical form for the $B$ matrix that remains the same
regardless of coordinate transformations. \ Note also that this becomes one in
the case of $f_{n}=f_{m}=1,$ which is expected. \ This completes our review of
the two-body formalism. \ Since we are attempting to reach a convergence point
with as few Gaussians as possible, we do not necessarily include as many
wavefunctions as is possible. \ So for $N=2$, we only begin with two
wavefunctions for each coordinate (giving a 4x4 matrix) and then go to three
wavefunctions (going to 9x9). \ Similarly, $N=3$ can have up to 5
wavefunctions per coordinate, but we only add one at a time in order to more
quickly converge the energy eigenvalues.

\end{document}